\documentclass[twocolumn,10pt]{infocom} 

\newcommand{\commentout}[1]{}

\include{epsf}

\newenvironment{ctable*}
{\begin{table*}[htpb]\begin{center}}{\end{center}\end{table*}}

\newtheorem{theorem}{Theorem}[section]

\newtheorem{corollary}[theorem]{Corollary}

\newtheorem{definitionx}[theorem]{Definition}

\begin{document}

\title{\Large\textbf Gossip-Based Ad Hoc Routing 
}

\author{ 
Zygmunt J. Haas
\hspace{10pt}  Joseph Y. Halpern \hspace{10pt}   Li Li%
\thanks{The work of Z.~Haas was supported in part by NSF under grant number
ANI-9980521 and ONR under contract number N00014-00-1-0564.
The work of J.~Halpern and L.~Li was supported in part by NSF under
grants grants IRI-96-25901, IIS-0090145, and NCR97-25251, and ONR under 
grants N00014-00-1-03-41, N00014-01-10-511, and N00014-01-1-0795.}
 \\
School of Electrical and Computer Engineering/Department of Computer Science \\
Cornell University \\
{haas@ece.cornell.edu} \hspace{10pt} \{halpern,lili\}@cs.cornell.edu \\
}

\date{}

\maketitle

\begin{abstract} 
Many {\em ad hoc\/} routing protocols are based on 
some variant of
flooding. Despite  various
optimizations, 
many routing messages are propagated unnecessarily.
We propose a gossiping-based approach,
where each node forwards a message with some probability,
to reduce the
overhead of the routing protocols.  
Gossiping exhibits bimodal behavior in sufficiently large networks: 
in some executions, the gossip dies out quickly and hardly any node gets
the message; in the remaining executions, a substantial fraction of the
nodes gets the message.
The fraction of executions in which
most nodes get the message depends on the gossiping probability
and the topology of the network.
In the networks we have considered, using gossiping probability 
between 0.6 and 0.8 suffices to ensure that almost every node gets the
message in almost 
every execution.
For large networks, this simple
gossiping protocol
uses 
up to 35\%
fewer messages than flooding, with improved performance.  
Gossiping can also be combined with various optimizations of flooding
to yield further benefits.
Simulations show that adding gossiping to AODV 
results
in significant performance improvement, 
even in networks
as small as 150 nodes.
We expect that the improvement
should be even more significant in larger networks.
\end{abstract}


\section{Introduction}
\label{sec-introduction}

An {\em ad hoc network\/}
%
is a multi-hop wireless network with no
fixed infrastructure. 
Rooftop networks and sensor networks are two existing 
types of
networks that
might be implemented using
the ad hoc networking technology.
Ad hoc networks 
can be
usefully
deployed in applications
such as disaster relief, tetherless classrooms, and battlefield
situations.

In ad hoc networks, the power supply of individual nodes is
limited, wireless bandwidth is limited, and the channel condition can
vary greatly.  Moreover, since nodes 
can be mobile, routes may
constantly change.  Thus, to enable efficient communication, 
robust routing protocols must be developed.

Many ad hoc routing protocols have been proposed.  Some, such as 
LAR~\cite{LAR98}, GPSR~\cite{GPSR00}, and DREAM~\cite{DREAM98} assume
that nodes are equipped with GPS hardware and thus know their
locations; others, such as DSR~\cite{DSR96}, AODV~\cite{AODV99},
ZRP~\cite{ZRP98}, and TORA~\cite{TORA97},  
do not make  this assumption.
Essentially all protocols that do not use GPS (and some that do, such as
LAR and DREAM) make use of flooding, usually with some optimizations.
%
\commentout{
Typically these are either  {\em structural
optimizations}, which impose  structure on the underlying network, or
{\em locality optimizations}, which take advantage of locality
information.  Among protocols that use structural optimizations, DCA
imposes a structural hierarchy on the underlying network through
clustering; OLSA
imposes structure through special multipoint relays; ZRP
calculates zones around each node and uses the zone information to
improve routing. The locality optimization can be temporal, spatial
or topological. 
Temporal locality is based on the notion that a given route will be up
for some time, even though nodes move around.
For example, the caching policy in DSR and AODV
exploits
temporal locality. 
Spatial locality exploits the assumption that 
if a node is close to a destination, it will continue to be close to
the destination for a some period of time.
Local repair 
(that is, the idea of finding a ``patch'' to a route if one link is broken)
in AODV exploits spatial locality. 
In \cite{Das99}, what might be called {\em topological locality\/} is
used.
The idea is that if a path from $u$ to $v$ no longer works, then a new
path will not differ much from the original path.
}

Despite the optimizations,  
in routing protocols that use flooding,
many routing messages are propagated unnecessarily.
In this paper, we
show that {\em gossiping\/}---essentially, tossing a coin to decide
whether or not to forward a message---can be used 
to significantly reduce the number of routing
messages sent.

%
It follows from results in percolation theory
\cite{Grimmett89,meester96} that gossiping exhibits a certain type of
bimodal behavior.   
Let the gossip probability be $p$.  
Then, in sufficiently large ``nice'' graphs, 
there are fractions $\theta^S(p)$ and
$\theta^R(p)$ such that 
the gossip quickly dies out in  
$1-\theta^S(p)$ of the executions and, in almost all of the fraction
$\theta^S(p)$ of the executions where the gossip does not die out, a
fraction $\theta^R(p)$ of the nodes get the message.  Moreover, in many
cases of interest, $\theta^R(p)$ is close to 1.  Thus, in almost all
executions of the algorithm, either hardly any nodes receive the
message, or most of them do.  
Ideally, we could make the
fraction of executions where the gossip dies out relatively low
while also keeping the gossip probability low, to reduce the message overhead.
The goal of this paper is to investigate the extent 
to which this can be done.  Our results show that, by using
appropriate heuristics,  
we can save 
up to 35\%
message overhead compared to flooding.
Furthermore, adding gossiping to a protocol such as 
AODV not only gives improvements in the number of messages sent, but
also results in improved network performance in terms of end-to-end latency
and throughput.  (For readers unfamiliar with AODV, a brief overview is given
  in Section\ref{sec-AODVoverview}.) 
We expect that the various 
optimizations applied to flooding
by other protocols 
(for example, the cluster-based scheme of \cite{NTCS99})
can also
be usefully combined with
gossiping to get further performance improvements.

We are certainly not the first to use gossiping in networking
applications.  For example, it has  been applied in networked
databases to spread updates among nodes \cite{Demers87} and 
to multicasting
\cite{Birman99}.   However, in almost all of the earlier work on
gossiping, it is assumed that any node in the network can send a
message to any other node, either because there is a direct link to
that node or a route to that node is known.  Gossiping proceeds by
choosing some set of nodes at random to which to gossip.
We do not have the luxury of being able 
to make such an assumption
in the context of
ad hoc networks.  Our problem is to 
{\em find routes\/}
to different nodes.

In an ad hoc network, if a message is transmitted by a node, 
due to the broadcasting 
nature of radio communications, the message is usually received by all
the nodes one hop away from the sender. Because of the fact that
wireless resources are expensive, it makes sense to take advantage of
this physical-layer broadcasting feature of the radio transmission.
In our gossiping protocol, we control the probability with which this
physical-layer broadcast is sent.

There has been some recent work on applying gossiping in ad hoc
networks, but the focus and thus the techniques used have been very different
from our work. 
Vahdat and Becker \cite{Vahdat00} apply gossiping 
to ad hoc unicast routing.
However, their usage of
gossiping is very different from ours.
In their work, they try to
ensure that messages are eventually delivered 
even if there is 
no connected path between the source and the
destination at any given point in time. As long as there exists a path
using communication links at some point in time, 
%
messages can be delivered through a random pair-wise exchanges among
mobile hosts. 
Their techniques are not intended for (and would not
perform well in) our setting,
 where we are trying to find routes that we assume exist
(because network partition is a rare event).
Chandra et al.~\cite{CRBirman01} 
use a 
gossiping mechanism 
to improve multicast reliability in ad hoc networks; 
they do not use gossiping to reduce the number of messages sent.
Indeed, they start with an arbitrary, possibly 
unreliable, multicast protocol to multicast a message.  They then
use gossiping (under the assumption that routes are known) to randomly
exchange messages between nodes in order to recover lost messages. 
Heinzelman et al.~\cite{Hari99} have applied
gossiping in data dissemination in wireless sensor networks,  
using techniques similar in spirit to those of \cite{Vahdat00}.
Again, the setting and results are quite different from ours.
Ni et al.~\cite{NTCS99} propose five different approaches to reduce
broadcast redundancy.  One of them (briefly mentioned in a few
sentences) is 
gossiping.  
However, they do not study the properties of gossiping nor do they
consider heuristics for dealing with problems introduced by gossiping 
in realistic ad hoc network topologies.  Their experiments do show,
however, that,  
in a 100-node network, using gossiping can save messages.

The rest of this paper is organized as follows:
Section~\ref{sec-thresh} 
discusses the basic bimodal effect in more detail.
Section~\ref{sec-bimodalexp} provides experimental evidence of the
bimodal effect in networks of reasonable size, and also gives a sense of
how the probability varies with the average degree of the network and
initial conditions. Section~\ref{sec-gossip} presents a number of
heuristics that should improve the performance of gossiping in networks
of interests, and investigates the extent to which they do so
experimentally.  
Section~\ref{sec-aodv} shows that gossiping can help in practical
settings, by considering the effect of adding gossiping to AODV.
We show by simulation that even in networks with 150 nodes only, adding
gossiping to AODV can result in significant performance improvements on
all standard metrics.  We expect that this improvement will be even more
significant in larger networks.
Section~\ref{sec-conclude} concludes our paper.

\section{The bimodal behavior of gossiping}
\label{sec-thresh}
Since flooding is a basic element
in many of the ad hoc routing protocols, as mentioned in Section 
\ref{sec-introduction}, we start by comparing gossiping to flooding.

Our basic gossiping protocol is simple. A source sends the route
request with probability 1. When a node first receives a route
request, with probability $p$ it broadcasts the request to its
neighbors and with probability $1-p$ it discards the request; if  the
node receives the same route request again, it is discarded.
Thus, a node broadcasts a given route request at most
once.   
This simple protocol is called GOSSIP1($p$). 

GOSSIP1 has a slight problem with initial conditions.  If the source has
relatively few neighbors, there is a chance that none of them will
gossip, and the gossip will die.  To make sure this does not happen, we
gossip with probability 1 for the first $k$ hops 
before continuing to
gossip with probability $p$.  We call this modified protocol
GOSSIP1($p,k$).%
\footnote{Of course, the fact that gossiping has difficulties if a node
has relatively few neighbors is true not just 
initially.  We return
to this point in the next section when we discuss optimizations.}

The performance of GOSSIP1($p,k$) clearly depends on the choice of $p$
and $k$.  Clearly, GOSSIP1(1,1) is equivalent to flooding.  
What happens in general?  
That depends in part on the 
topology of the network (particularly the average degree of the network nodes), 
the gossip probability $p$, and the initial conditions (as
determined by $k$).  If we think of gossiping 
as spreading a disease in an epidemic,
this simply says that the likelihood of an epidemic
spreading depends in part on how many people each person can infect
(the degree), the likelihood of the infection spreading (the gossip
probability), and how many people are initially infected.  

As we said in the introduction, gossiping and, in particular,
the performance of GOSSIP1($p$,0) (that is, the scenario where even the
source gossips with probability $p$) has been well studied in the
work on percolation theory \cite{Grimmett89,meester96}.  Quite a few
types of networks have been studied in the literature. 
In this section, we focus on two of them.
We first study regular networks, since they allow us to easily analyze how 
GOSSIP1 behaves with respect to different parameters, such as the gossip
probability, network 
size, and node degree, without other complicating factors. We then study
random networks constructed as follows. 
Nodes are placed at random on a two-dimensional area; 
an edge is placed between any pair of nodes less than a fixed
distance $d$ apart.  This type of random graph seems appropriate for
modeling a number of applications involving ad hoc networks.   Nodes
have a limited amount of transmission power, and so can communicate only with
reasonably close nodes.  The random placement can be viewed as modeling
features such as the random mobility of nodes and the random placement
of sensors 
in a large region.

The following theorem gives a sense of the type of results that have
been proved. 
\begin{theorem}
\label{theorem-1}
For all $p \ge 0$, for all infinite regular graphs $G$, and for 
almost all (i.e., a measure 1 subset) of the infinite random graphs $G$
constructed as above, 
if GOSSIP1($p$,0) is used by every node to spread a
message, 
%
%
then there is a well-defined probability 
$\theta^S_0(p)<1$ 
that the message
reaches infinitely many nodes. 
Moreover, the probability 
$\theta^F_0(p)$
that a node
receives the message and forwards it in an execution where the message
reaches infinitely many nodes 
is equal to $\theta^S_0(p)$.%
\footnote{Note that our bimodal effect is different from~\cite{Birman99}. They
  describe a bimodal behavior where either all or no process
  receives the multicast message.  
}
\end{theorem}

Note that the probability of a message dying out (i.e., not
spreading to infinitely many nodes) is averaged over the executions
of the algorithm.  That is, the theorem says that if we execute the
algorithm repeatedly, the probability that a message does not die out in
any given execution is 
$\theta^S_0(p)$.  
On the other hand, 
$\theta^F_0(p)$
talks about the probability that a node receives and
forwards the message in a given execution of the algorithm.  
The intuition behind the equality of 
$\theta^S_0(p)$ and $\theta^F_0(p)$
is easy to explain. A gossip initiated by a source $n_0$ dies out if
there is a set of nodes $N$ that 
disconnects $n_0$ from the rest of the graph; that is, every infinite
path starting at $n_0$ must go through a node in $N$.  Thus, 
$\theta^S_0(p)$ 
is the probability that there is no disconnecting set $N$ such that none
of the nodes in $N$ forward the message.  (Note that $N$ could consist
of the singleton node $n_0$ itself.)  Similarly, the probability 
$\theta^F_0(p)$
that a
random node $n$ 
receives and forwards the message is precisely
the probability that there is 
no  set $N'$ such that $N'$ disconnects $n$
from $n_0$
and none of the nodes in $N'$ forwards the message.
%
Therefore, $\theta^S_0(p)=\theta^F_0(p)=_{\mbox{\small def}}\theta_0(p)$.

It follows from these results that, in an execution where the
message does not die out, the probability that a random node receives
the message is $\theta_0(p)/p$, since receiving the message is
independent of forwarding it.  Thus, in terms of the notation used in
the introduction, $\theta^S(p) = \theta_0(p)$ and $\theta^R(p) =
\theta_0(p)/p$.  

Let $\theta^S_k(p)$ be the probability that a message reaches infinitely
many nodes if GOSSIP1($p,k$) is used.  It is easy to see that
$\theta^S_1(p) = \theta_0(p)/p$, since the probability that the message
reaches infinitely many nodes using GOSSIP1($p,1$) is precisely the
probability that a message reaches infinitely many nodes using
GOSSIP1($p,0$) given that the source actually gossips.  However, note
that the probability that a node receives and forwards a message if
GOSSIP1($p,k$) is used, given that the message does not die out, is
still $\theta_0(p)$.  
That is, the probability that a node receives the message is independent
of the choice of $k$.
On the other hand, it is not hard to see that if each node learns the
network topology in a zone of radius $k$ (so that it can route a message
directly to any node in its zone), then the probability that a node
receives and forwards a message given that the message does not die out
is $\theta_k(p)$. 

All these results are for infinite graphs.  It is not hard to
show that essentially the same results hold for finite graphs, except
possibly near the boundary.  
In sufficiently large finite
graphs, there will be two types of executions: those where hardly any
node gets the message, and those where the message makes it all the way
to the boundary.  It follows easily from the Central Limit Theorem that,
in sufficiently large graphs, in almost all executions where the gossip
does not die out, a fraction $\theta_0(p)/p$ nodes will get the message.
That is, we expect the bimodal behavior: either hardly any nodes get the
message, or a fraction $\theta_0(p)/p$ receive the message.  As we shall
see, in cases of interest, $\theta_0(p)$ is quite close to $p$.  Thus,
in almost all executions of the algorithm in sufficiently large graphs,
either hardly any nodes receive the message, or most do.  

This leads to a number of obvious questions:
\begin{itemize}
\item How large is ``sufficiently large''?
\item What is the behavior of $\theta_k(p)$ for different graphs
of interest?
\item What can be done to improve the performance of gossiping in
realistic settings?
\end{itemize}
We investigate these questions in the next two sections.

\section{Gossiping in finite networks}
\label{sec-bimodalexp}
We did a number of experiments to investigate the behavior of gossiping.
We summarize some of the more interesting results
here.  
\commentout{
Our experimental results suggest that, in 
sufficiently large graphs, either the epidemic dies out very quickly or
almost everyone gets the disease.  More precisely, we show that, in
almost every execution of the gossip protocol, either almost every node
gets the message or hardly any node does.  The fraction of
executions in which almost every node gets the message depends on
factors like the gossip probability and the network topology.  
If the gossip probability is above a certain threshold, then almost
every node 
gets the message in almost every execution of the protocol.
To demonstrate the threshold effect, we experimented on a number of
networks.
}
We 
assumed
an ideal MAC layer
for these experiments because we wanted to decouple the effect of the
MAC layer from the effect of gossiping.   
An ideal MAC layer is one that is not 
subject to packet loss.
When we consider more realistic scenarios in Section \ref{sec-aodv}, 
we use the IEEE 802.11 MAC layer.
\commentout{ 
We first study regular networks, since they allow us to easily analyze how 
GOSSIP1 behaves with respect to different parameters, such as the gossip
probability, network size, and node degree, without other complicating
factors. As we shall see, the behavior in regular graphs
seems quite indicative of the behavior of gossiping in general. We
then study random networks constructed as follows:
Nodes are placed at random on a two-dimensional area; 
an edge is placed between any pair of nodes less than a fixed
distance $d$ apart.  This type of random graph seems appropriate for
modeling a number of applications involving ad hoc networks.   Nodes
have a limited amount of power, and so can communicate only with
reasonably close nodes.  The random placement can be viewed as modeling
features such as the random mobility of nodes and the random placement
of sensors 
in a large region.
} 
In this section, we focus on regular graphs and the random graphs
discussed in the previous section. 

\begin{figure}[ht]
\setlength\tabcolsep{0.1pt}
\begin{center}
\begin{tabular}{cc}
\epsfysize=4.0cm \epsffile{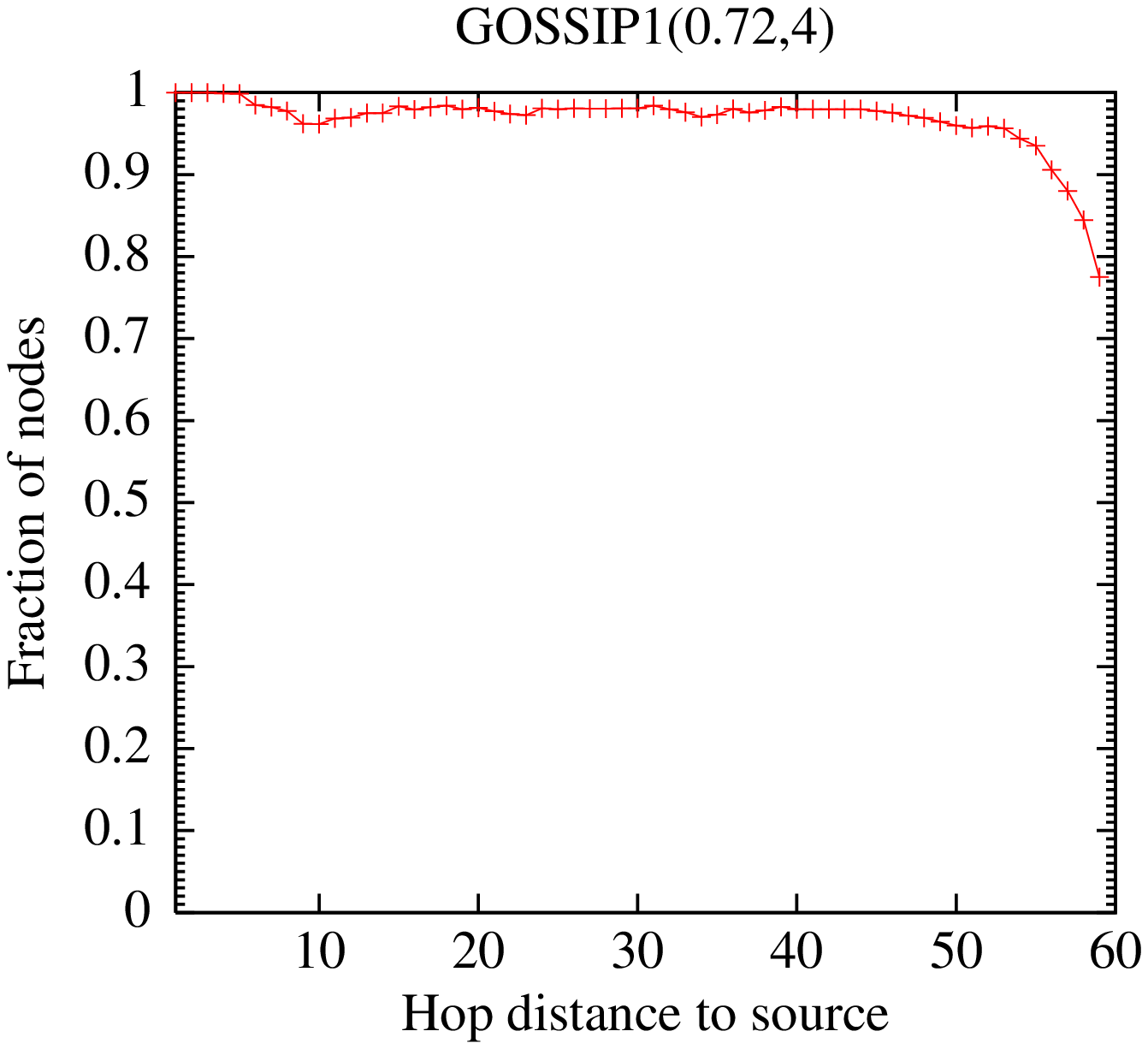}  & 
\epsfysize=4.0cm \epsffile{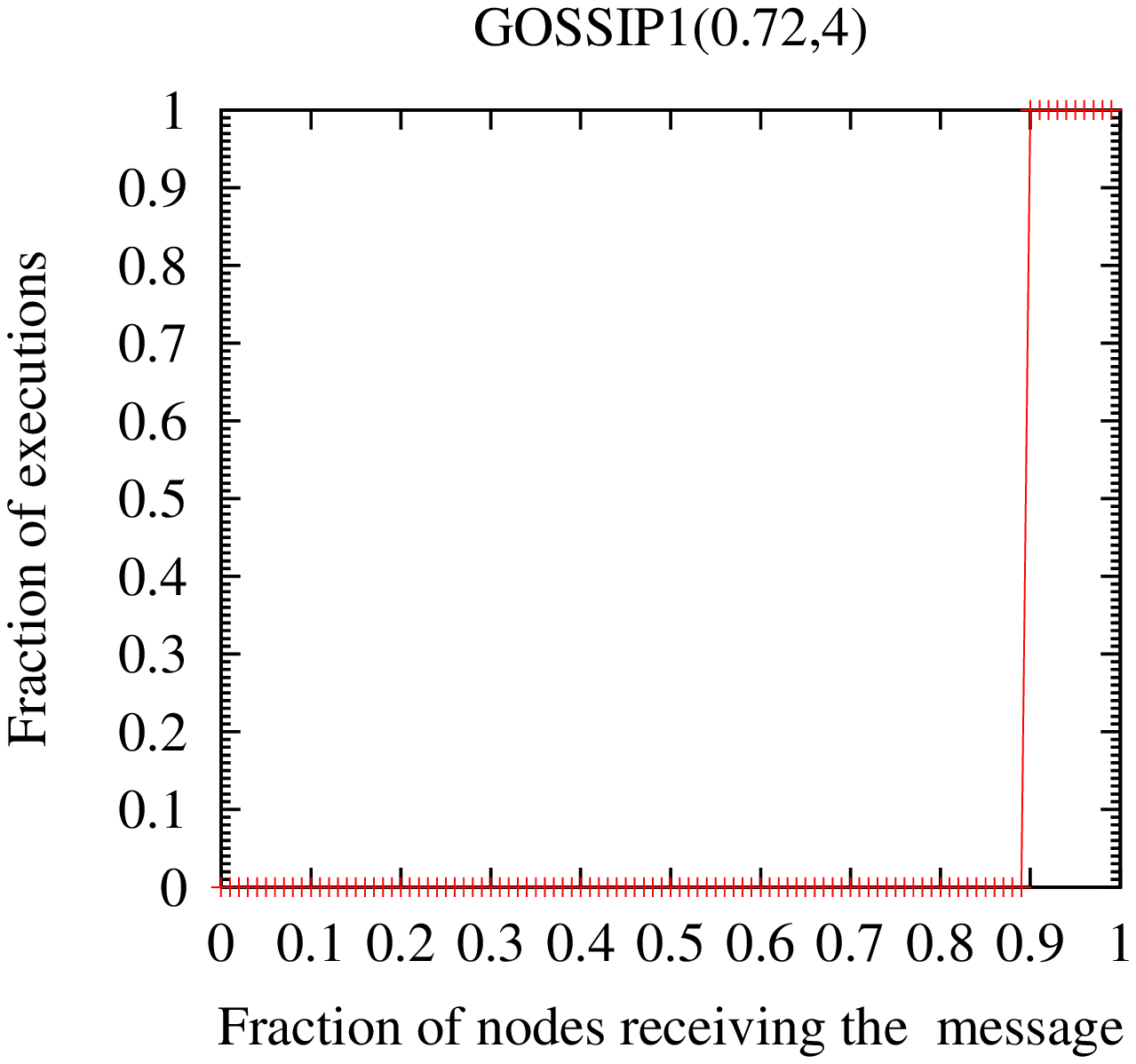} \\
{\footnotesize (a)} &
{\footnotesize (b)}  \\
\epsfysize=4.0cm \epsffile{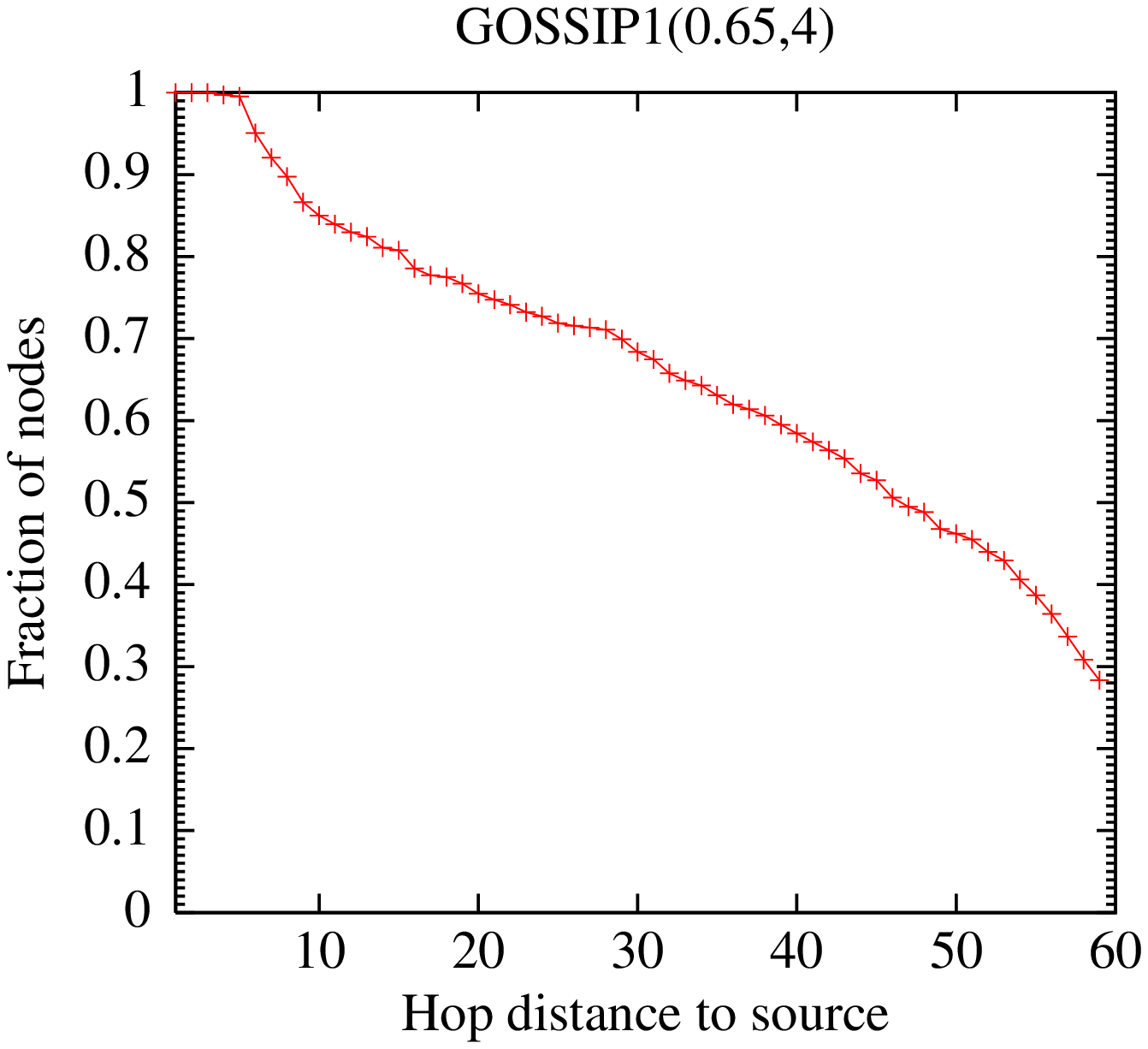}  &
\epsfysize=4.0cm \epsffile{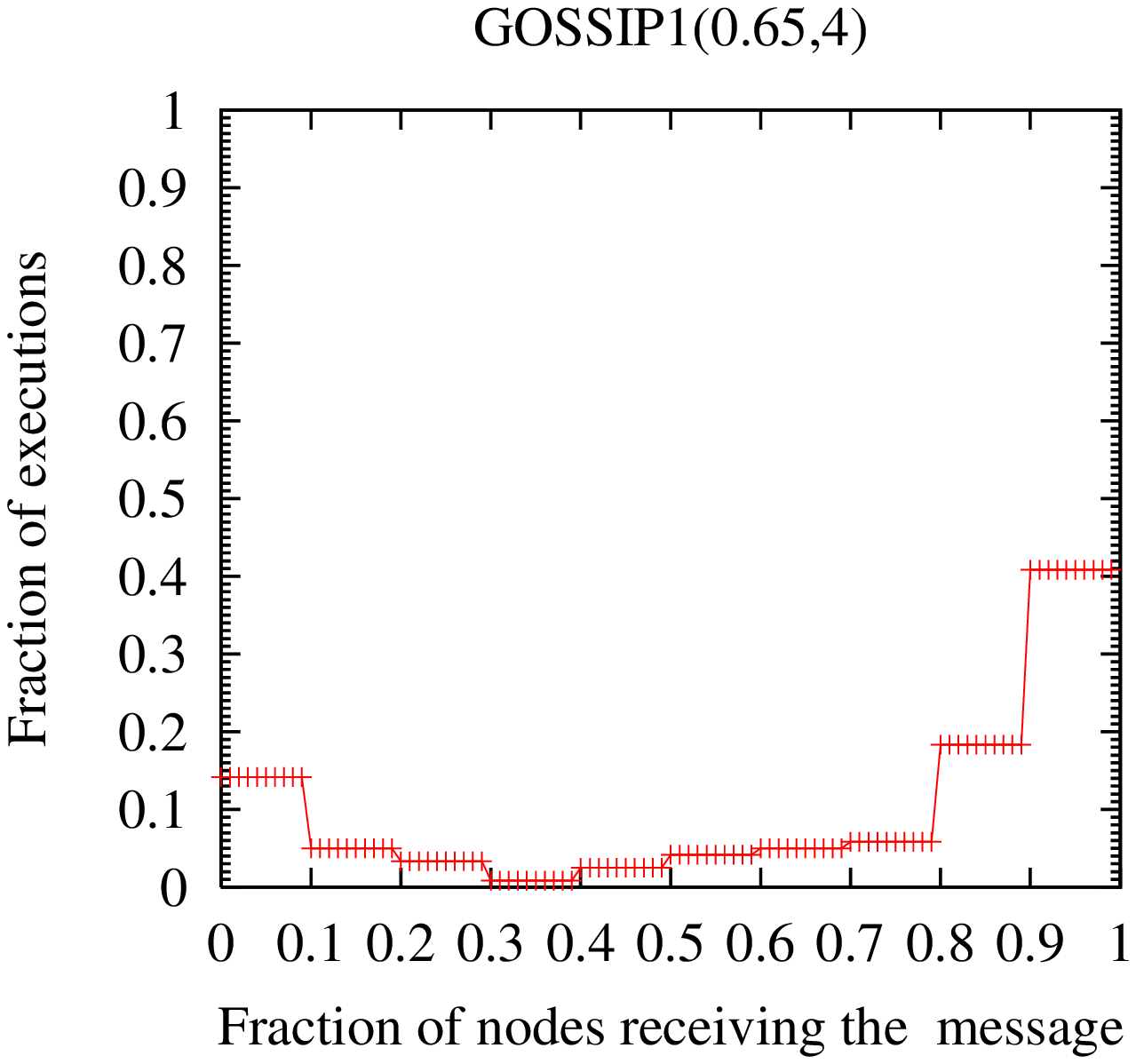} \\
{\footnotesize (c)} &
{\footnotesize (d)}  \\
\epsfysize=4.0cm \epsffile{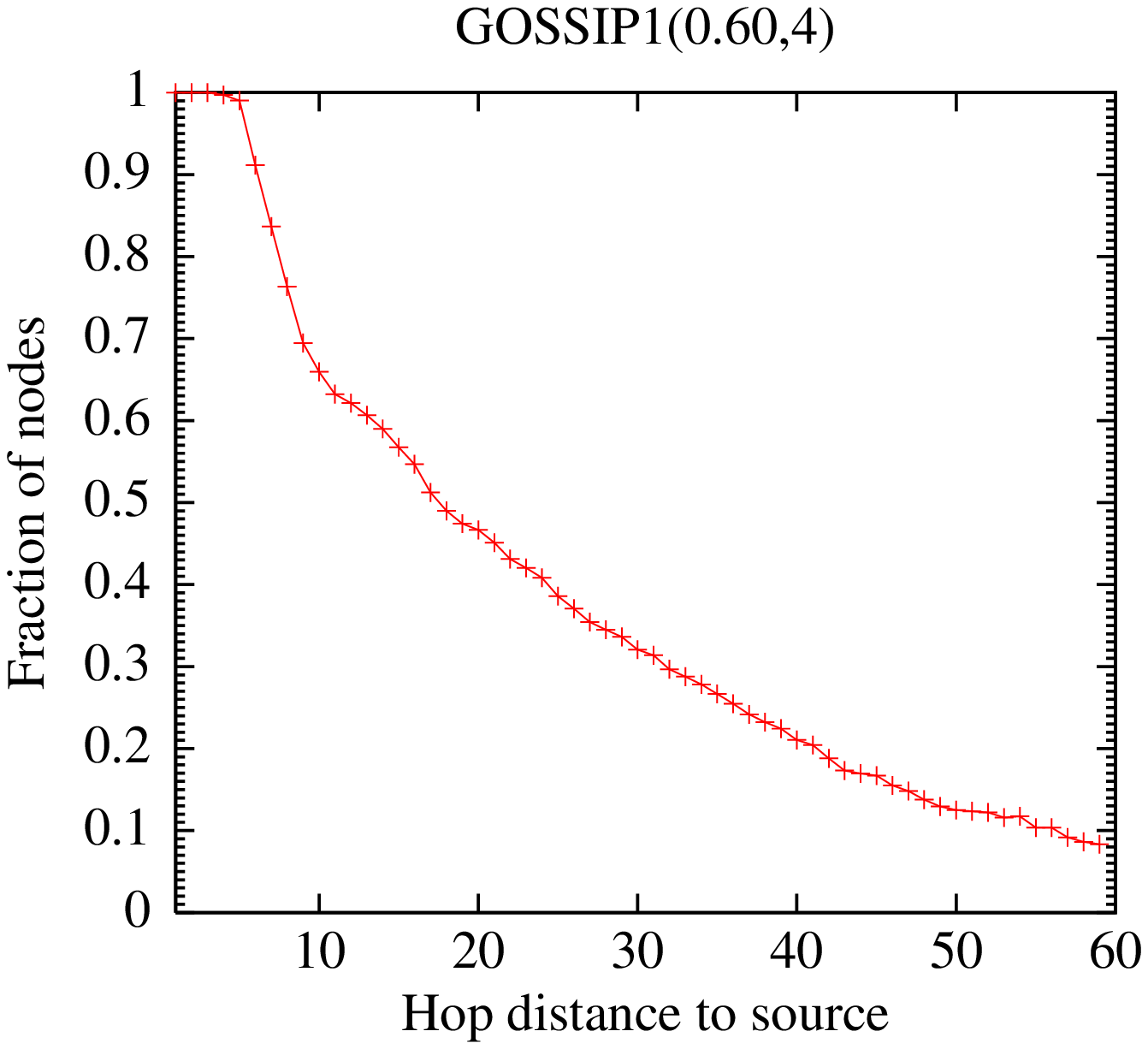}  & 
\epsfysize=4.0cm \epsffile{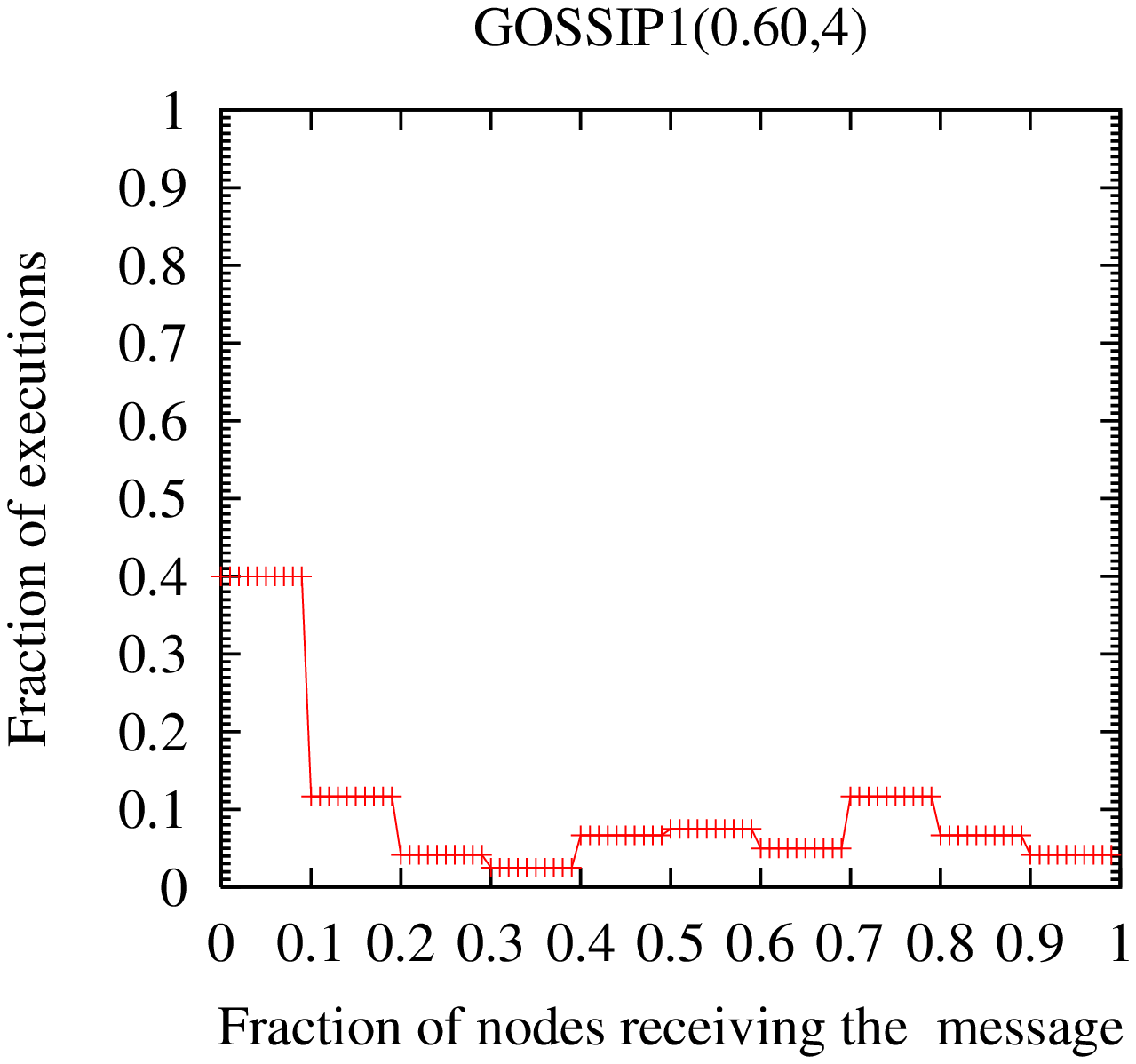} \\
{\footnotesize (e)} &
{\footnotesize (f)} 
\end{tabular}
\end{center}
\caption{The behavior of gossiping on a $20 \times 50$ grid.
\label{fig-grid}
}
\end{figure}

Our first set of experiments involves ``medium-sized'' networks, with
1000 nodes.  
We start by considering a 20-row by 50-column grid (i.e., a regular graph
of degree 4).  
We focus on GOSSIP1($p,4$), since taking $k=4$ produces a
reasonable tradeoff.  
(We report the effect of varying $k$ towards the end of this section.)
The results depend in part on where we place the route request source.
As we would expect from the theoretical arguments, the location of 
the source node does not affect the fraction of nodes
receiving the message.  However, it does affect the number of  
executions in which the gossip dies out.  
The number of executions in which the gossip does not die out is higher
for a more central node, and lower for a corner node.  
We report results here for the case where the route request source is at
the left boundary  
of row 10.  
Our experiments show that, on average, the performance 
for other locations of the route request source
is somewhat 
better than the results reported here.
The results are illustrated in Figure~\ref{fig-grid}.
Notice that GOSSIP1(.72,4) on the grid ensures that almost all nodes get
the message, except for a slight dropoff at distance greater than 50.
This dropoff is a boundary effect, which we discuss in more detail below.
Note that the graph in Figure~\ref{fig-grid}(a) represents an average
of 120 executions of the protocol.  
With gossip probability .72 for this grid size, in almost all the
executions of the algorithm, almost all nodes get the message.

\input{epsf}
\commentout{ 
\input{epsf}
\begin{figure}[htb]
\setlength\tabcolsep{0.1pt}
\begin{center}
\begin{tabular}{cc}
\epsfysize=4.0cm \epsffile{mobiFigures/hisGk4gridn1000p65.ps} &
\epsfysize=4.0cm \epsffile{mobiFigures/hisGk4gridn1000p60.ps} \\
{\footnotesize (a)} &
{\footnotesize (b)}  \\
\end{tabular}
\begin{tabular}{c}
\epsfysize=4.0cm \epsffile{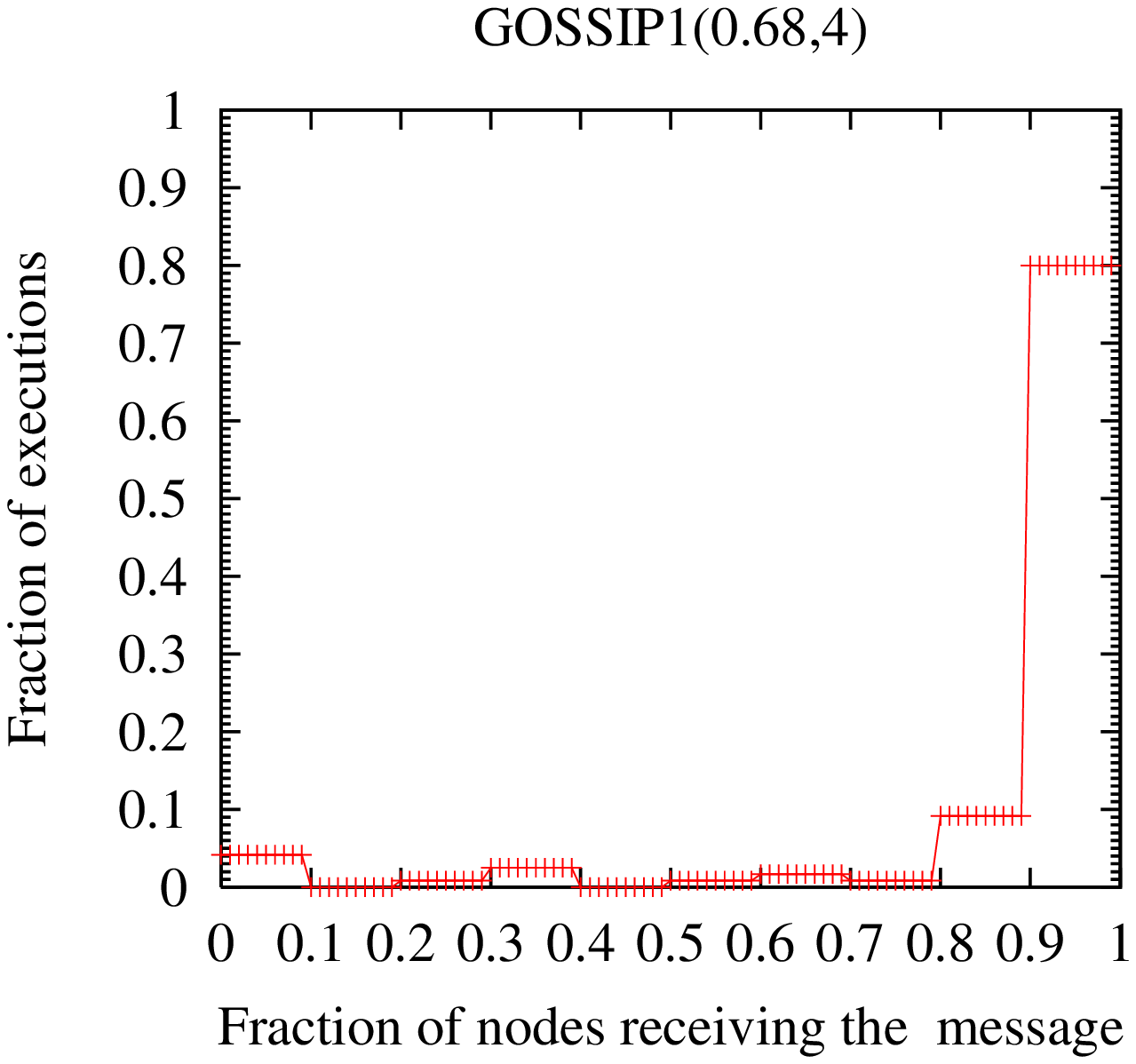}  \\
{\footnotesize (c)} 
\end{tabular}
\end{center}
\caption{The bimodal behavior of gossiping on a $20\times 50$ grid.
\label{fig-begrid}
}
\end{figure}
}
The situation changes significantly if the gossip probability is even
a little less than .7.  For example, the average performance of
GOSSIP1(.65,4) is shown in Figure~\ref{fig-grid}(c).  
As the graph
shows,  
at distance 40, on average 58\% of the nodes got
the message.  
However, in this case, the graph is somewhat misleading.
The averaging is hiding the true behavior.
As we would expect from Theorem~\ref{theorem-1}, there is bimodal behavior.
This is illustrated in Figure~\ref{fig-grid}(d).   
If we consider nodes at distance 15--45 (so as to
ignore initial effects and boundary effects), in 14\% of the executions,
fewer than 10\% of the nodes get the message; in 19\% of the executions, 
fewer than 20\% of the nodes get the message; in 59\% of the executions,
more than 80\% of the nodes get the message; and in 41\% of the
executions, more than 90\% of the nodes get the message.

If we lower the gossip probability further, we get the same bimodal
behavior; all that changes is the fraction of executions in which 
all nodes and no nodes get the message.  The dropoff is fairly rapid.  
For example, Figure~\ref{fig-grid}(e) and (f) describe
the situation for GOSSIP1(.60,4).  
By the time we get to probability .6 on the grid, in only 4\%
executions of the algorithm is it the case that 
more than 90\% of the nodes get the message; in only 11\% of the  executions
do more than 80\% of the nodes get the message; and in over 50\% of
the executions, fewer than 20\% of the nodes get the messages. 

We also investigated the effect of the degree of the network on
gossiping.  
Not surprisingly, increasing the degree makes it better and decreasing
it makes it worse.  
In a $20 \times 50$ regular network of degree 6,
it suffices
to gossip with 
probability .65 to ensure that almost all nodes get the message in
almost all executions; with gossip probability .6, we start to see
some dropoff.  
(Again, the numbers given in the graph are actually the result of
averaging over a number of executions of the algorithm, and mask the
bimodal behavior observed in the executions.) 
\commentout{ 
\input{epsf}
\begin{figure}[ht]
\setlength\tabcolsep{0.1pt}
\begin{center}
\begin{tabular}{cc}
\epsfysize=4.0cm \epsffile{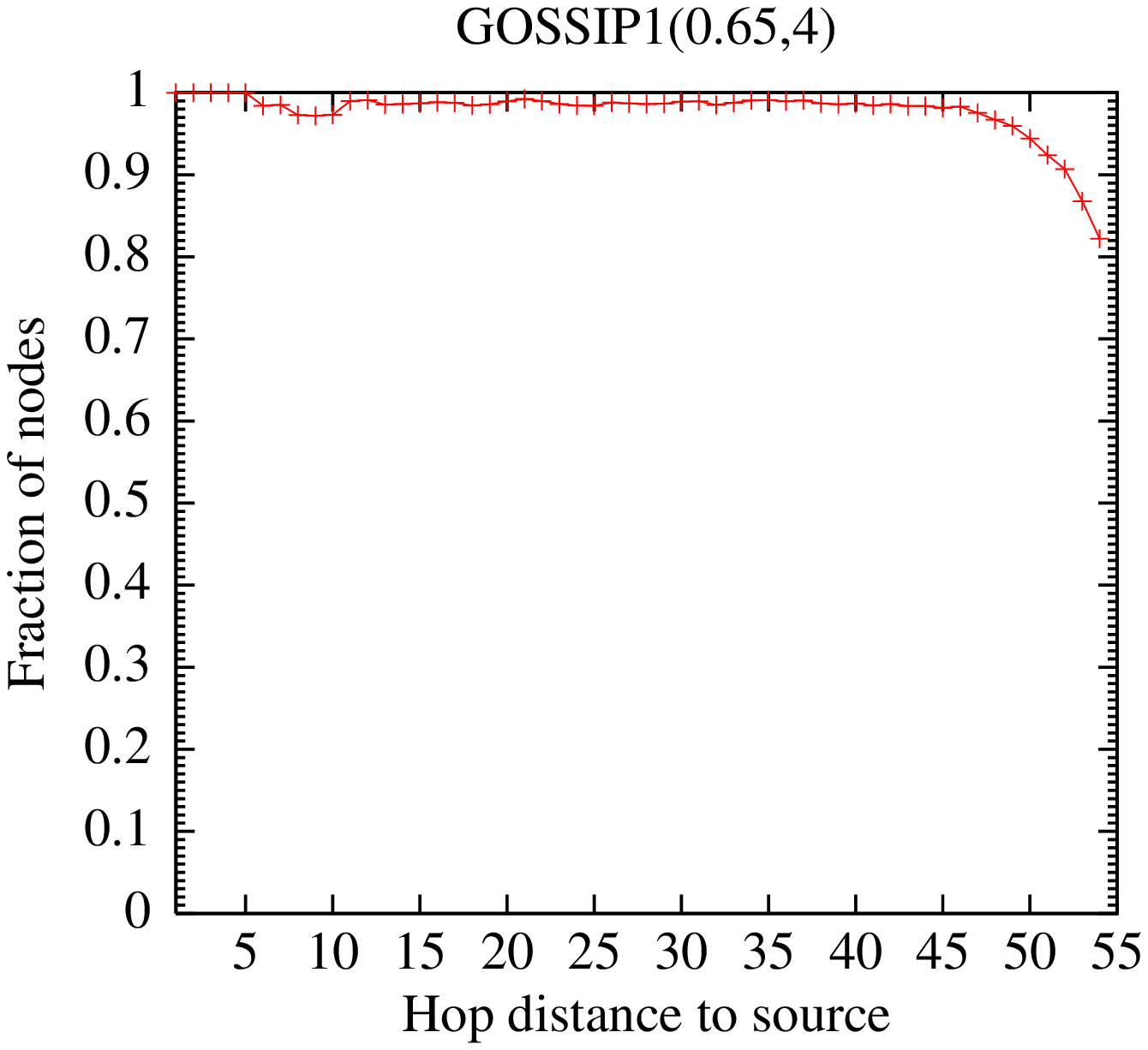}  & 
\epsfysize=4.0cm \epsffile{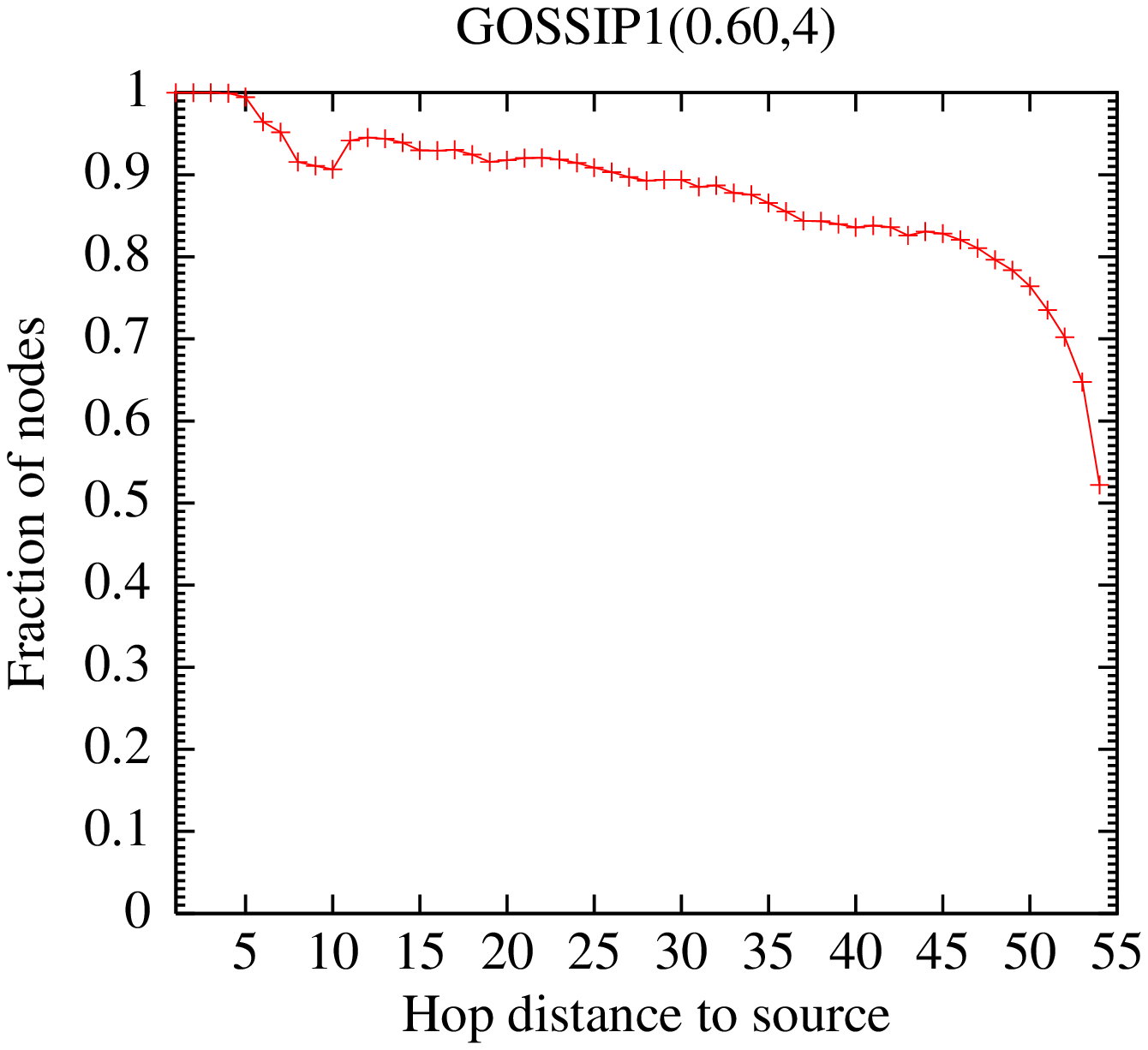}   \\
{\footnotesize (a)} &
{\footnotesize (b)}  
\end{tabular}
\end{center}
\caption{Gossiping on regular graph of degree 6.
\label{fig-reg6}
}
\end{figure}
} 
\noindent On the other hand, 
for a $20 \times 50$ regular network of 
degree 3, we need to gossip with  probability .86 to ensure that
almost all nodes get the message in all executions.
\commentout{ 
\input{epsf}
\begin{figure}[htb]
\setlength\tabcolsep{0.1pt}
\begin{center}
\begin{tabular}{cc}
\epsfysize=4.0cm \epsffile{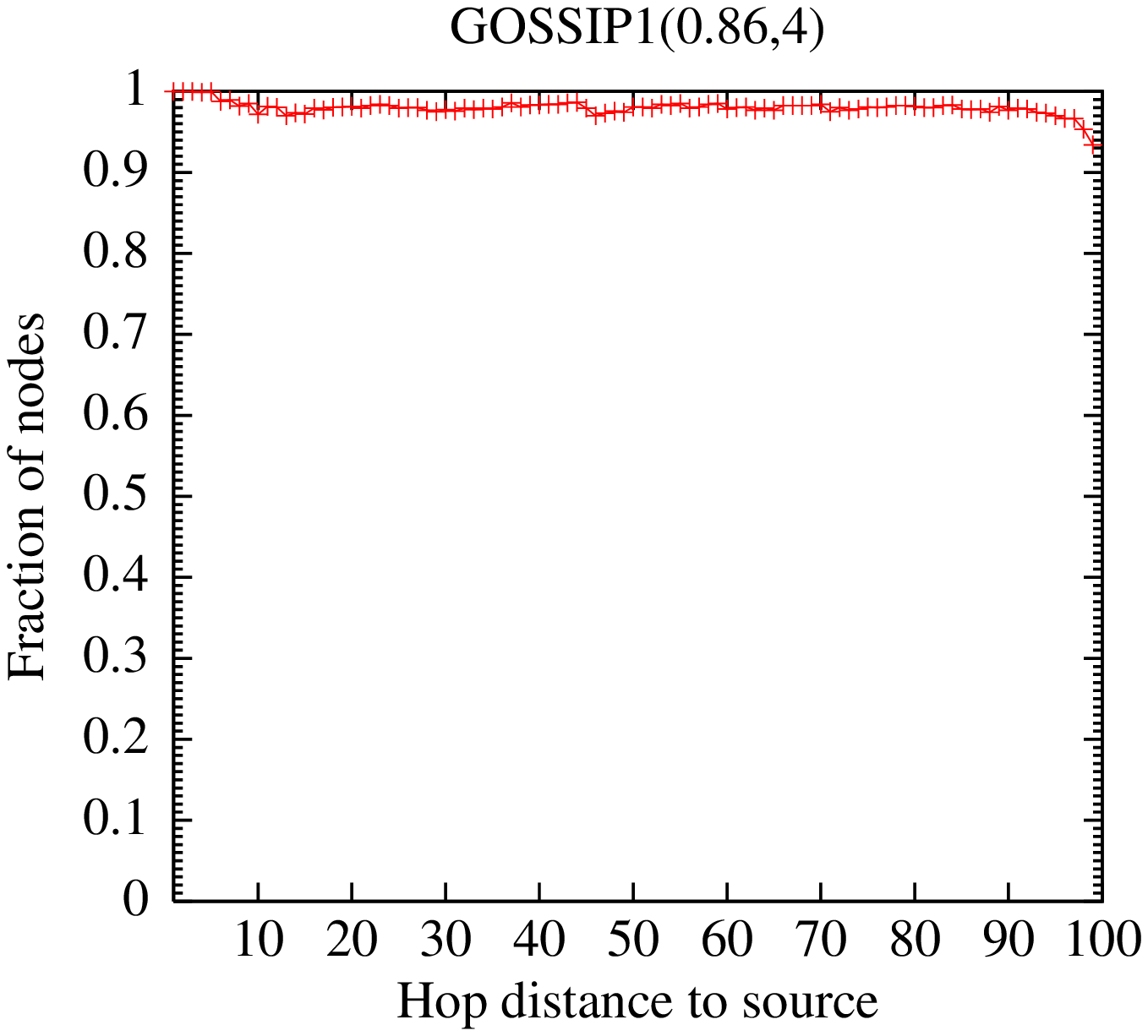}  & 
\epsfysize=4.0cm \epsffile{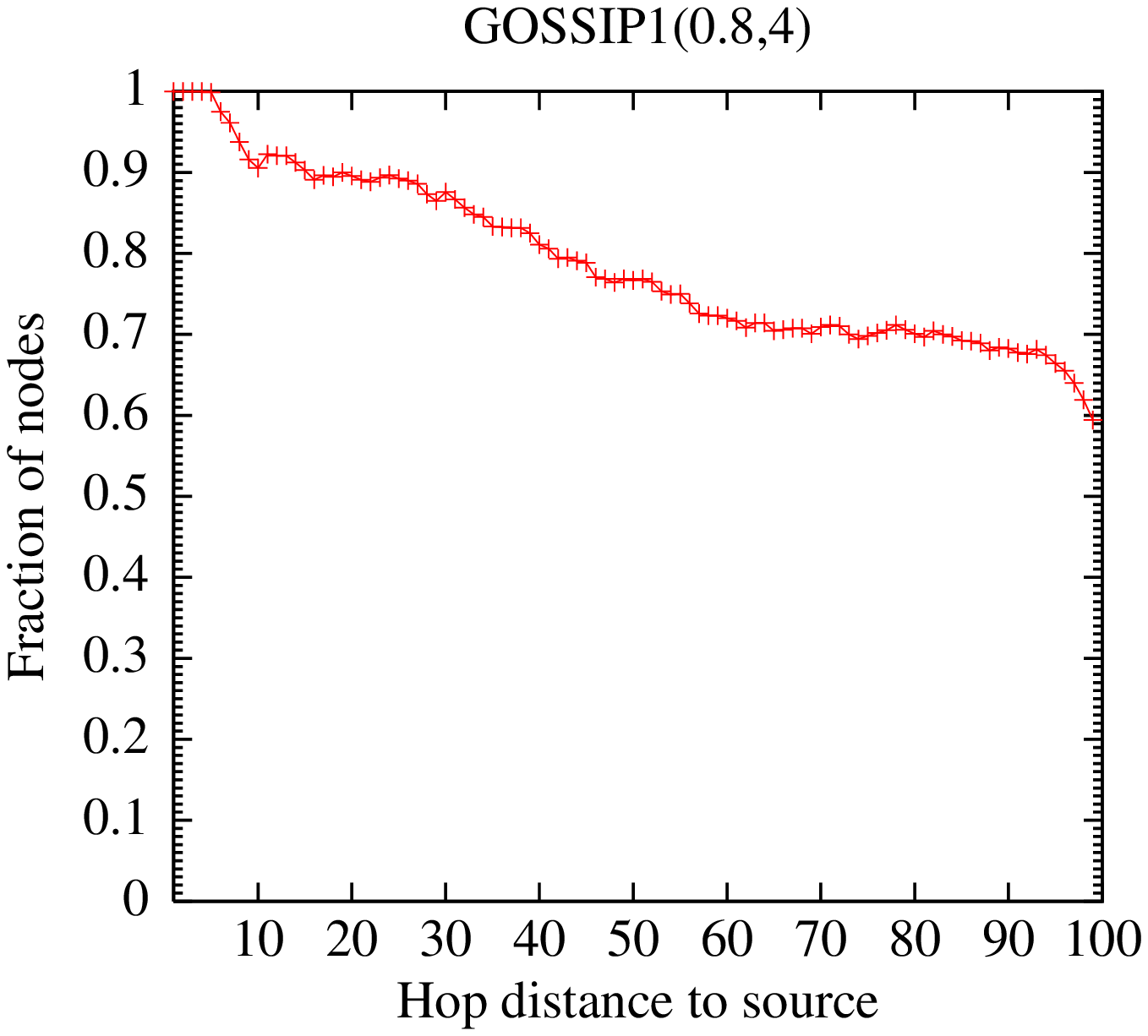}   \\
{\footnotesize (a)} &
{\footnotesize (b)}  
\end{tabular}
\end{center}
\caption{Gossiping on a regular graph of degree 3.
\label{fig-reg3}
}
\end{figure}
}

\begin{figure}[ht]
\setlength\tabcolsep{0.1pt}
\begin{center}
\begin{tabular}{cc}
\epsfysize=4.0cm \epsffile{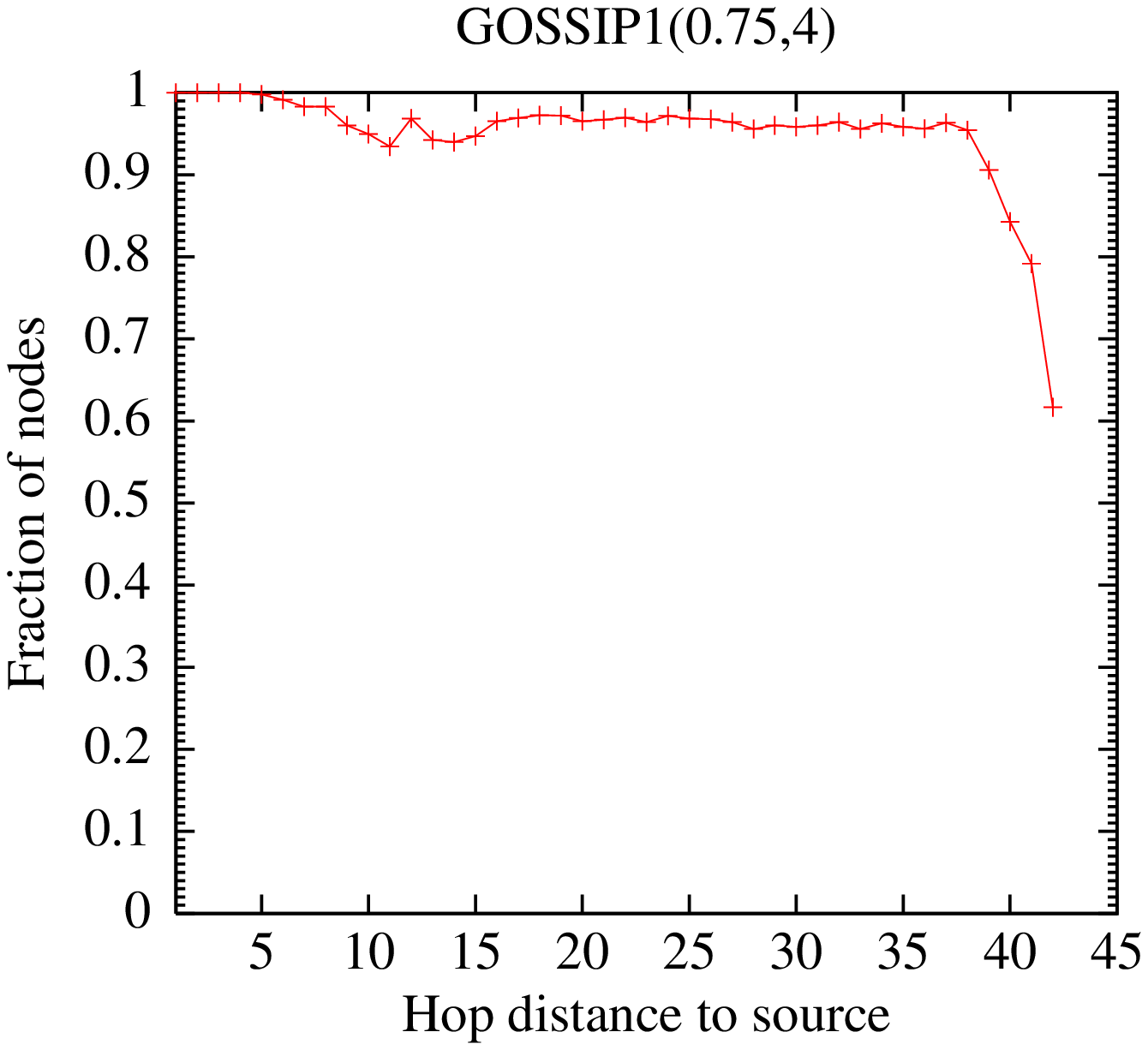}  &
\epsfysize=4.0cm \epsffile{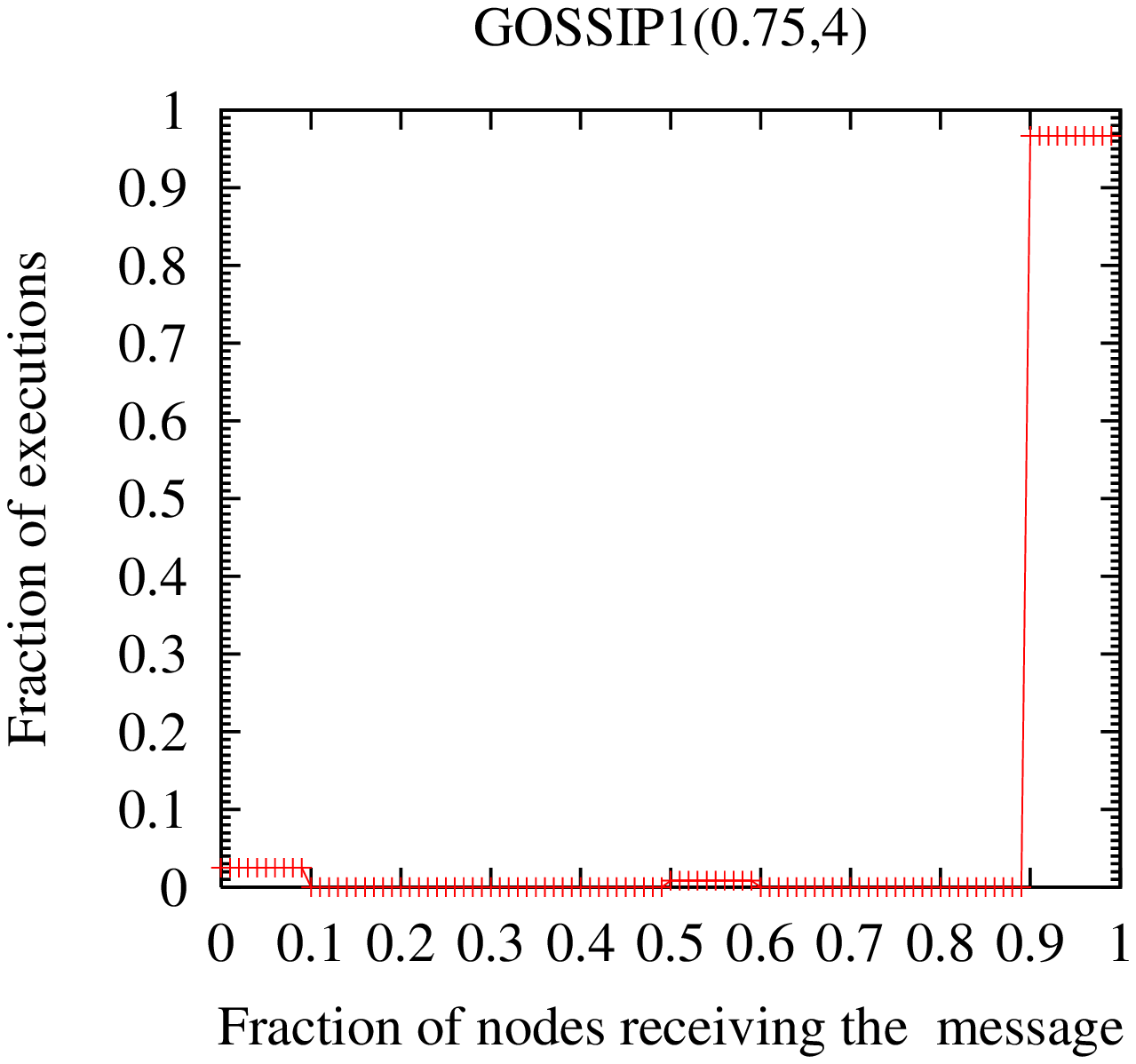} \\
{\footnotesize (a)} &
{\footnotesize (b)}  \\
\epsfysize=4.0cm \epsffile{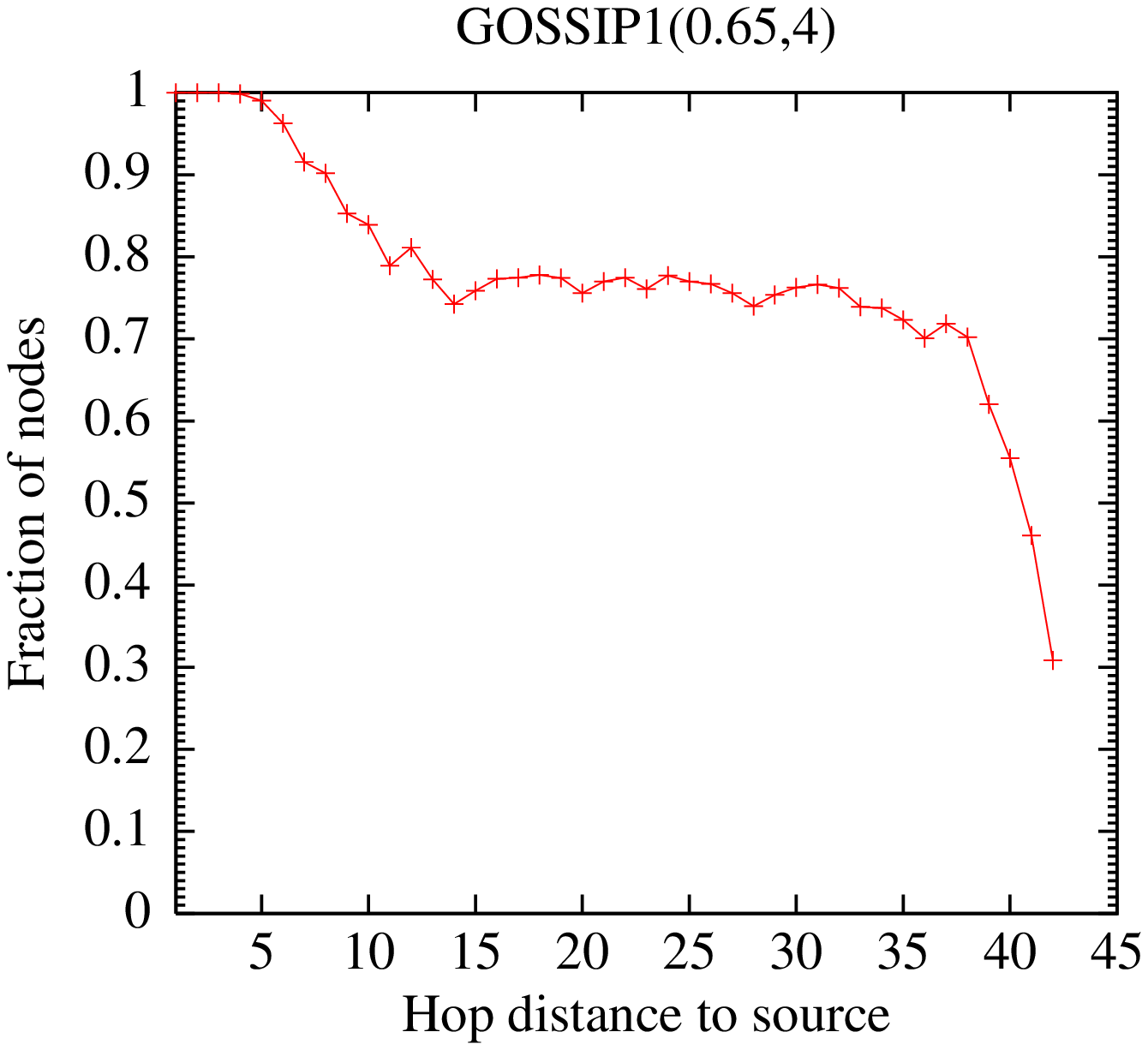} &
\epsfysize=4.0cm \epsffile{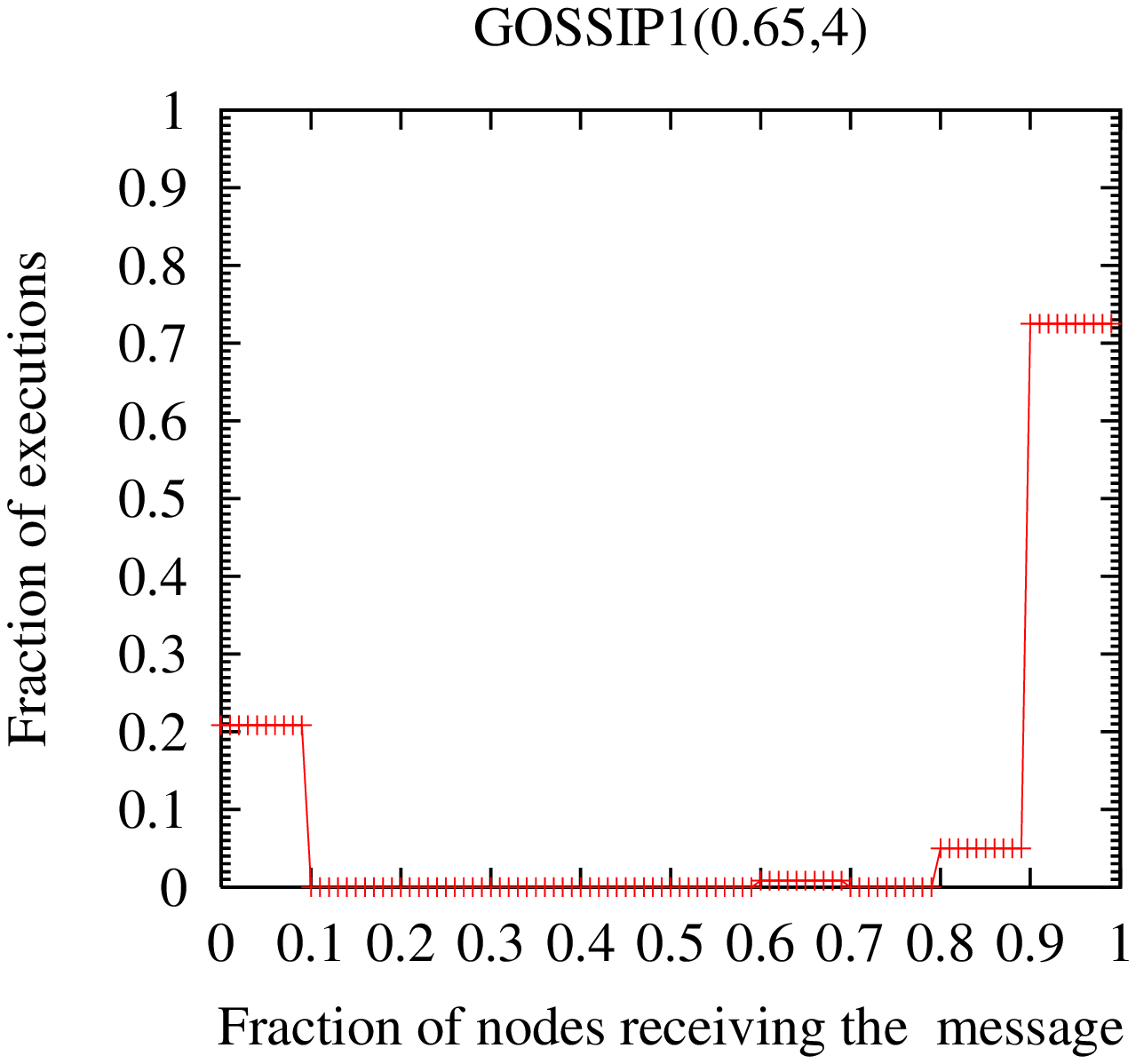} \\
{\footnotesize (c)} & 
{\footnotesize (d)} \\
\end{tabular}
\end{center}
\caption{Gossiping on a random network of average degree 8.
\label{fig-random8}
}
\end{figure}

While easy to study, regular graphs are not typical of the topology we
expect in practical ad hoc networks.  Random graphs are a somewhat better model.
We considered two families of random graphs.
In the first,
we randomly placed 1000 nodes 
in a $7500 m \times 3000 m$
rectangular region, where a node can communicate with another node if
it is no more than 250 meters away. This results in a network with average
degree 8.  
Since real networks have boundaries, we did not experiment on
wrap-around meshes.   As we shall see, dealing with nodes near the
boundary raises some interesting issues.  
The results 
of our experiments
are illustrated in Figure~\ref{fig-random8}.
\input{epsf}
The results are qualitatively similar to those on the grid, as we would
expect.  Indeed, 
the bimodal effect is particularly 
pronounced with GOSSIP1(.65,4), 
as shown in Figure~\ref{fig-random8}(d).
If we consider nodes at distance 15--35,
Figure~\ref{fig-random8}(d) shows,
in 20\% of the executions,
fewer than 10\% of the nodes get the message; in 70\% of the executions,
over 90\% of the nodes get the message, and in 75\% of the executions,
over 80\% of the nodes get the message.
\commentout{ 
\input{epsf}
\begin{figure}[htb]
\begin{center}
\begin{tabular}{c}
\epsfysize=4.0cm \epsffile{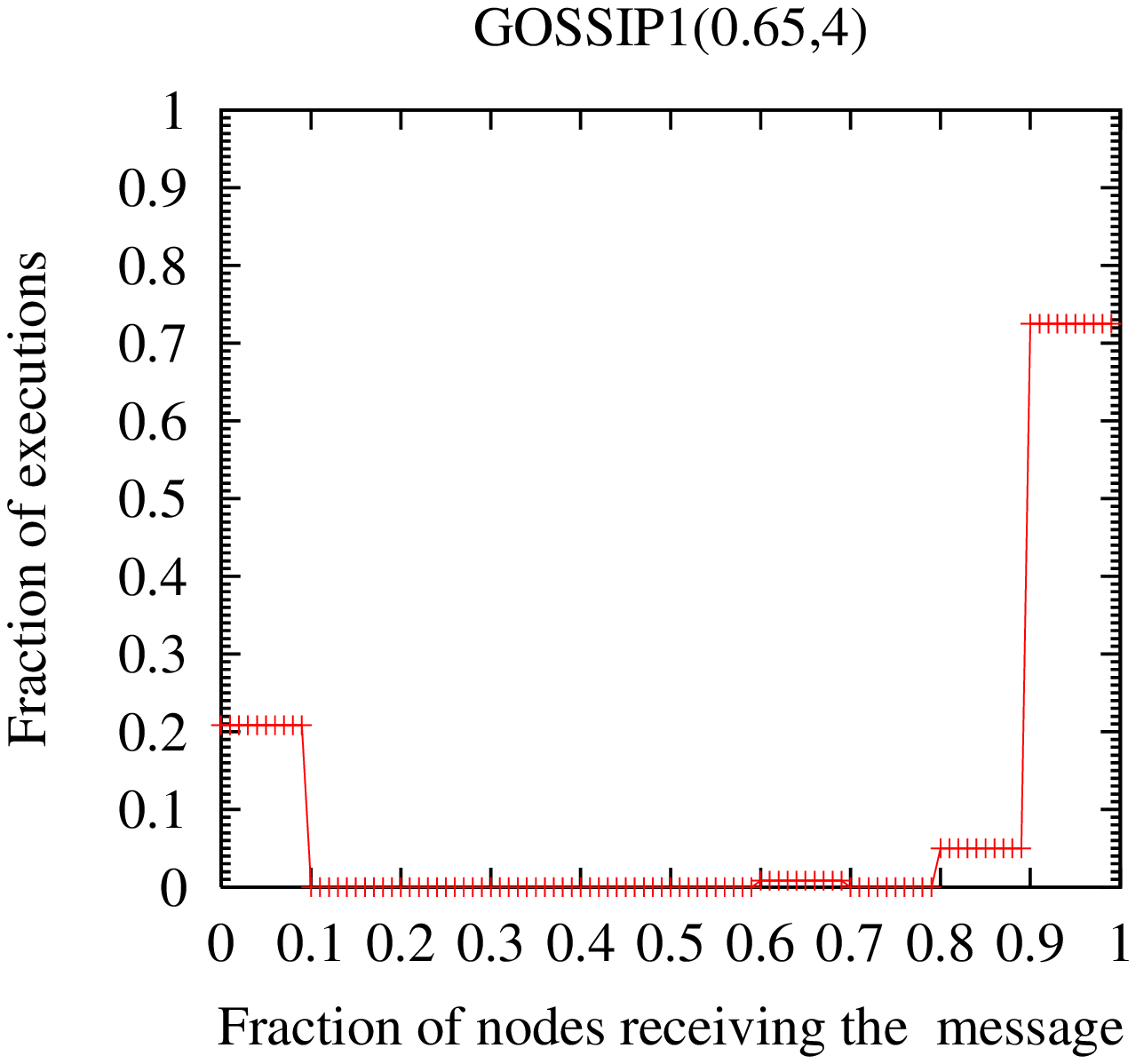} 
\end{tabular}
\end{center}
\caption{The bimodal behavior of gossiping on a random network of average degree 8.
\label{fig-berandom8}
}
\end{figure}
} 

To consider what happens with a higher-degree network,
we also placed 1200 nodes at random in the same rectangular
region; this results in a network with average degree 10. 
In this network, it suffices to gossip with probability .65 to ensure
that almost all nodes get the message in almost all executions. 
\commentout{ 
\input{epsf}
\begin{figure}[ht]
\setlength\tabcolsep{0.1pt}
\begin{center}
\begin{tabular}{cc}
\epsfysize=4.0cm \epsffile{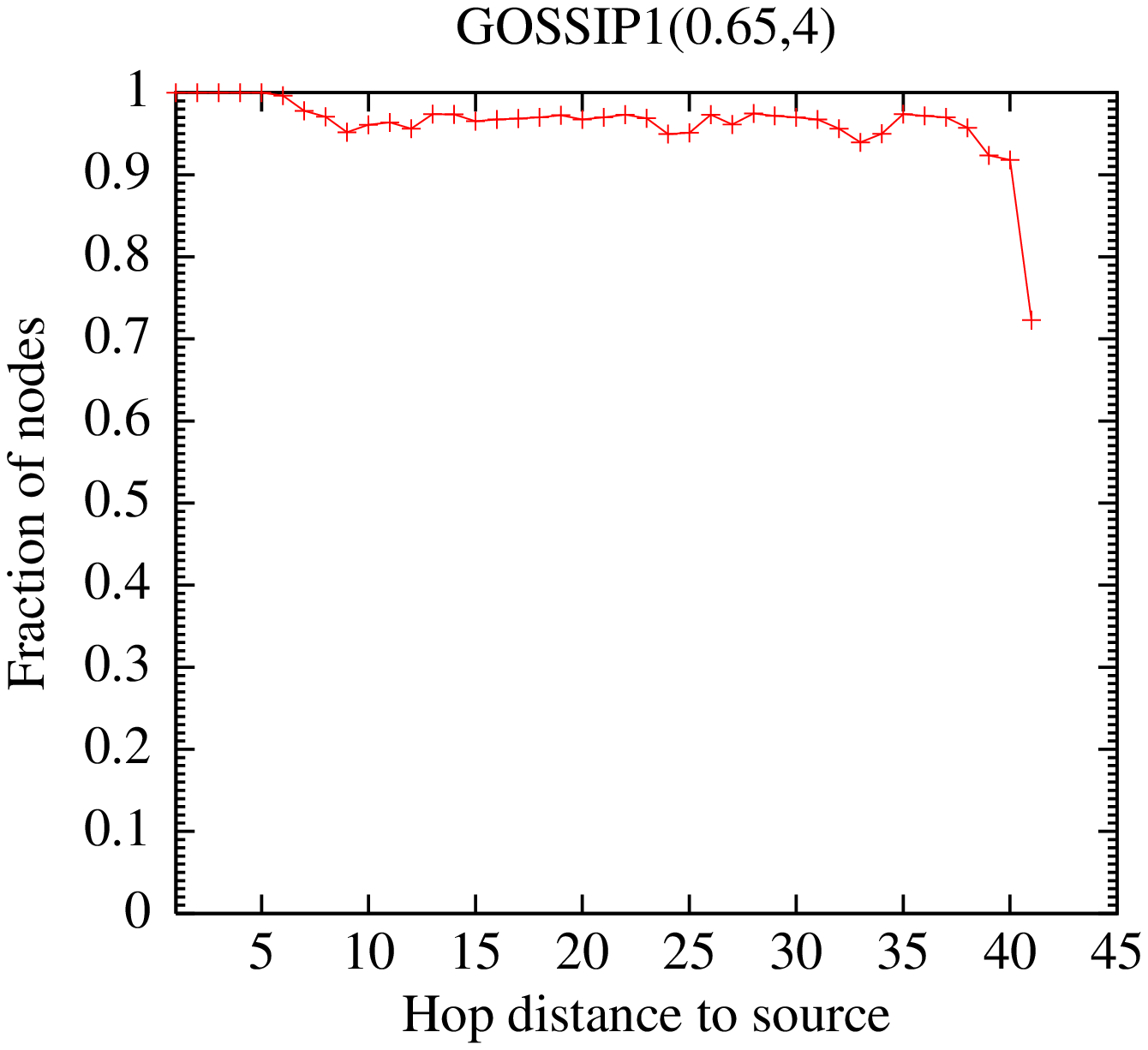}  &
\epsfysize=4.0cm \epsffile{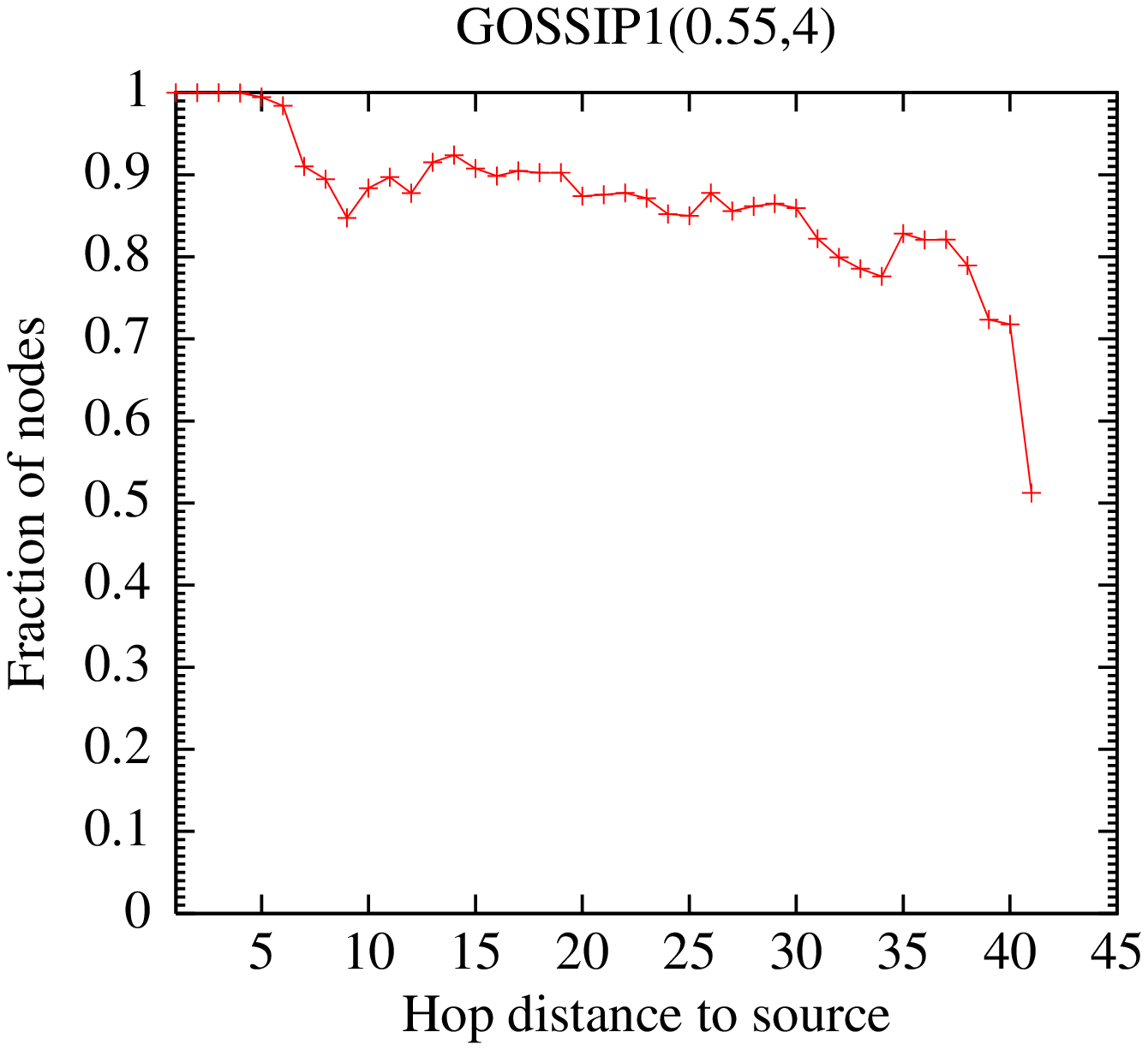} \\
{\footnotesize (a)} &
{\footnotesize (b)}  
\end{tabular}
\end{center}
\caption{Gossiping on a random network of average degree 10.
\label{fig-random10}
}
\end{figure}
}

All the graphs above show a marked
dropoff in 
probability for nodes that are close to the boundary.
This is not just an effect of averaging; this dropoff occurs in almost
all executions of the algorithm.  
The dropoff is due to two related boundary effects.
\begin{enumerate}
\item Distant nodes have fewer neighbors, since they are close to the
boundary.
\item Nodes at distance $d$ from the source may well receive
message due to ``back-propagation'' from nodes at distance $d' > d$
that get the message.  Such back-propagation is not possible for
boundary nodes.
\end{enumerate}
We discuss some techniques to deal with this dropoff 
in Section~\ref{zones}.

\begin{figure}[htb]
\setlength\tabcolsep{0.1pt}
\begin{center}
\begin{tabular}{cc}
\epsfysize=4.0cm \epsffile{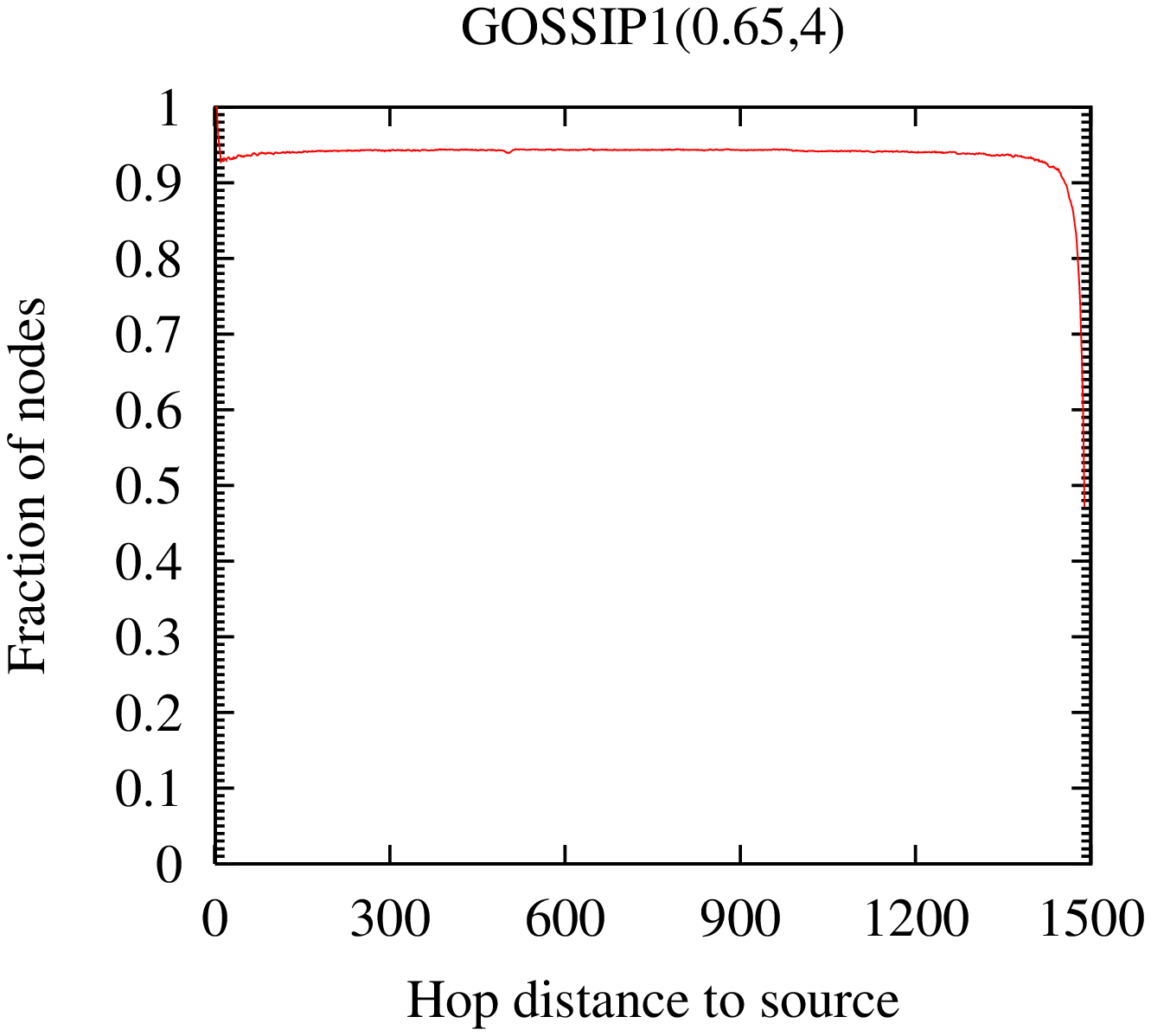}  & 
\epsfysize=4.0cm \epsffile{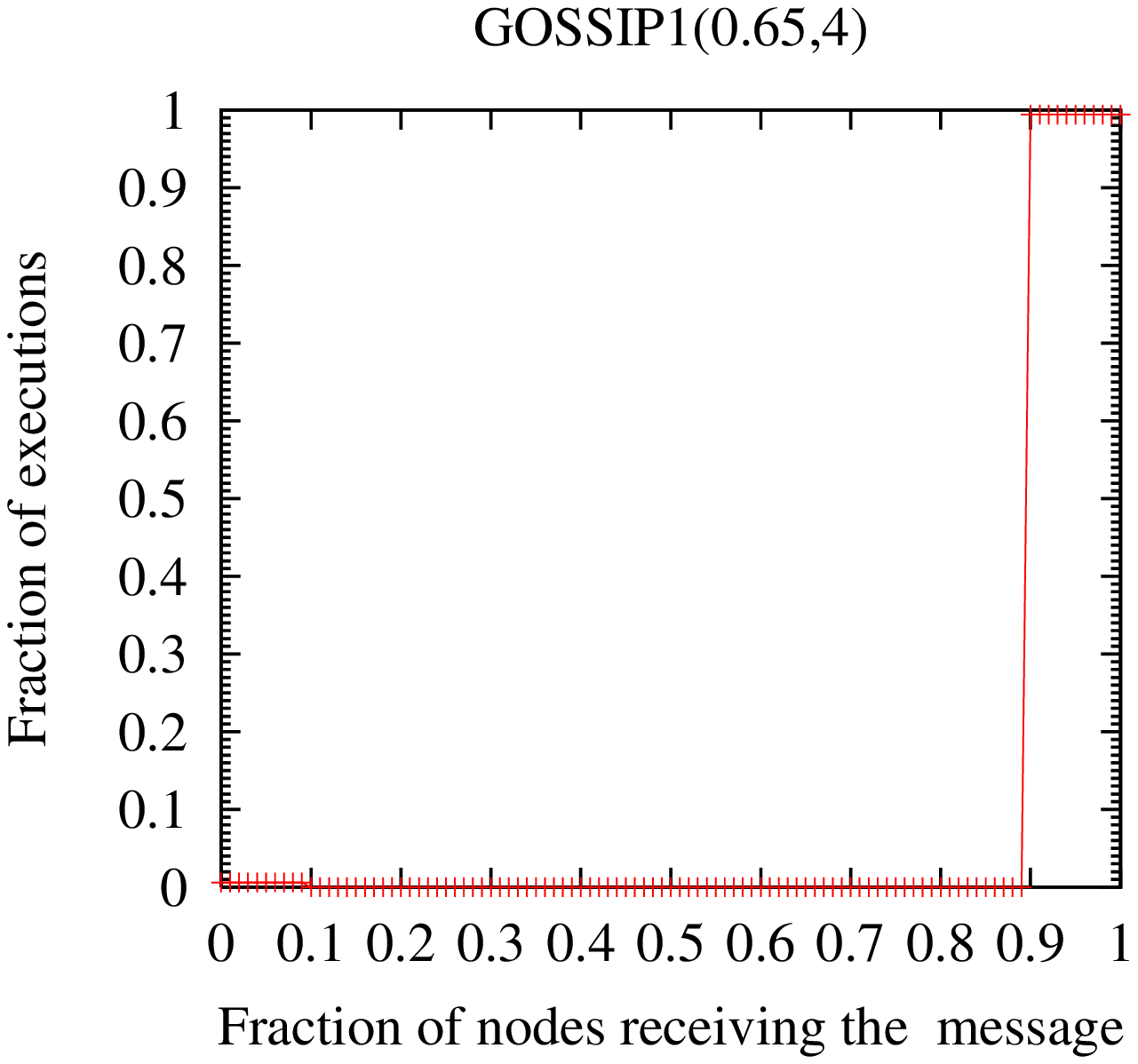}  \\
{\footnotesize (a)} &
{\footnotesize (b)}  \\
\epsfysize=4.0cm \epsffile{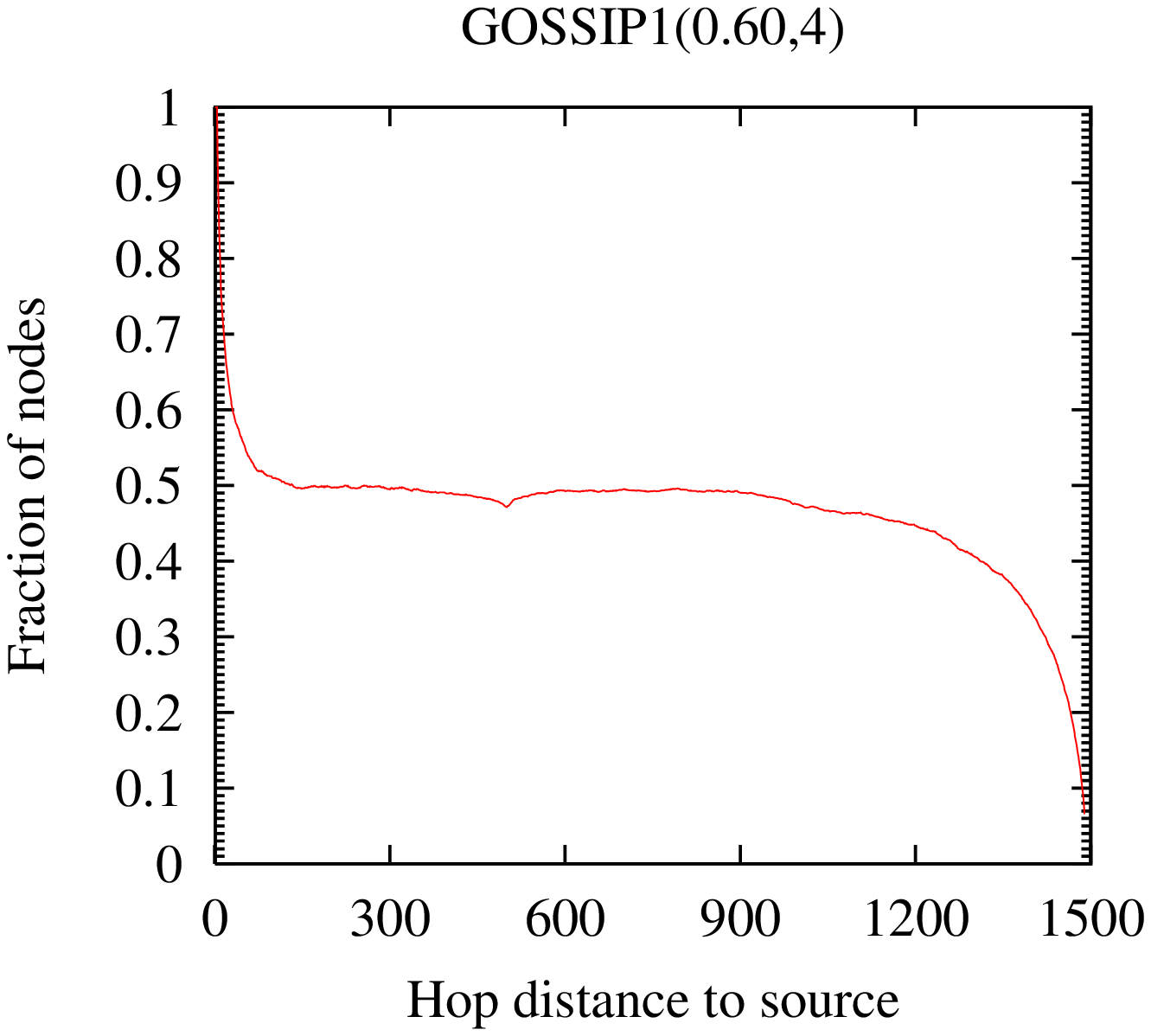} &   
\epsfysize=4.0cm \epsffile{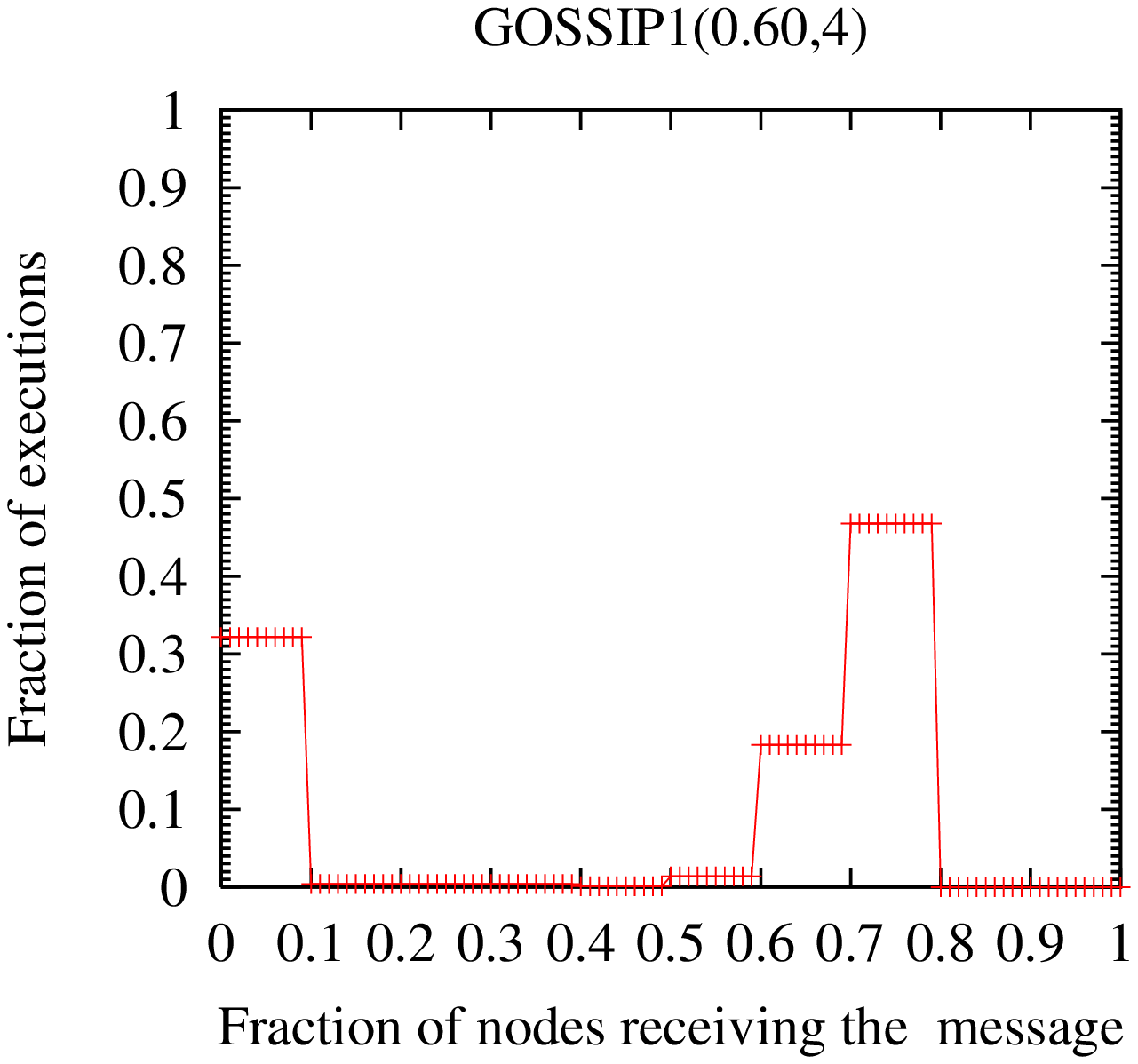}   \\
{\footnotesize (c)} &
{\footnotesize (d)} 
\end{tabular}
\end{center}
\caption{The behavior of gossiping on a 1000x1000 grid.
\label{fig-1mgrid}
}
\end{figure}

\begin{figure}[htb]
\setlength\tabcolsep{0.1pt}
\begin{center}
\begin{tabular}{c}
\epsfysize=4.0cm \epsffile{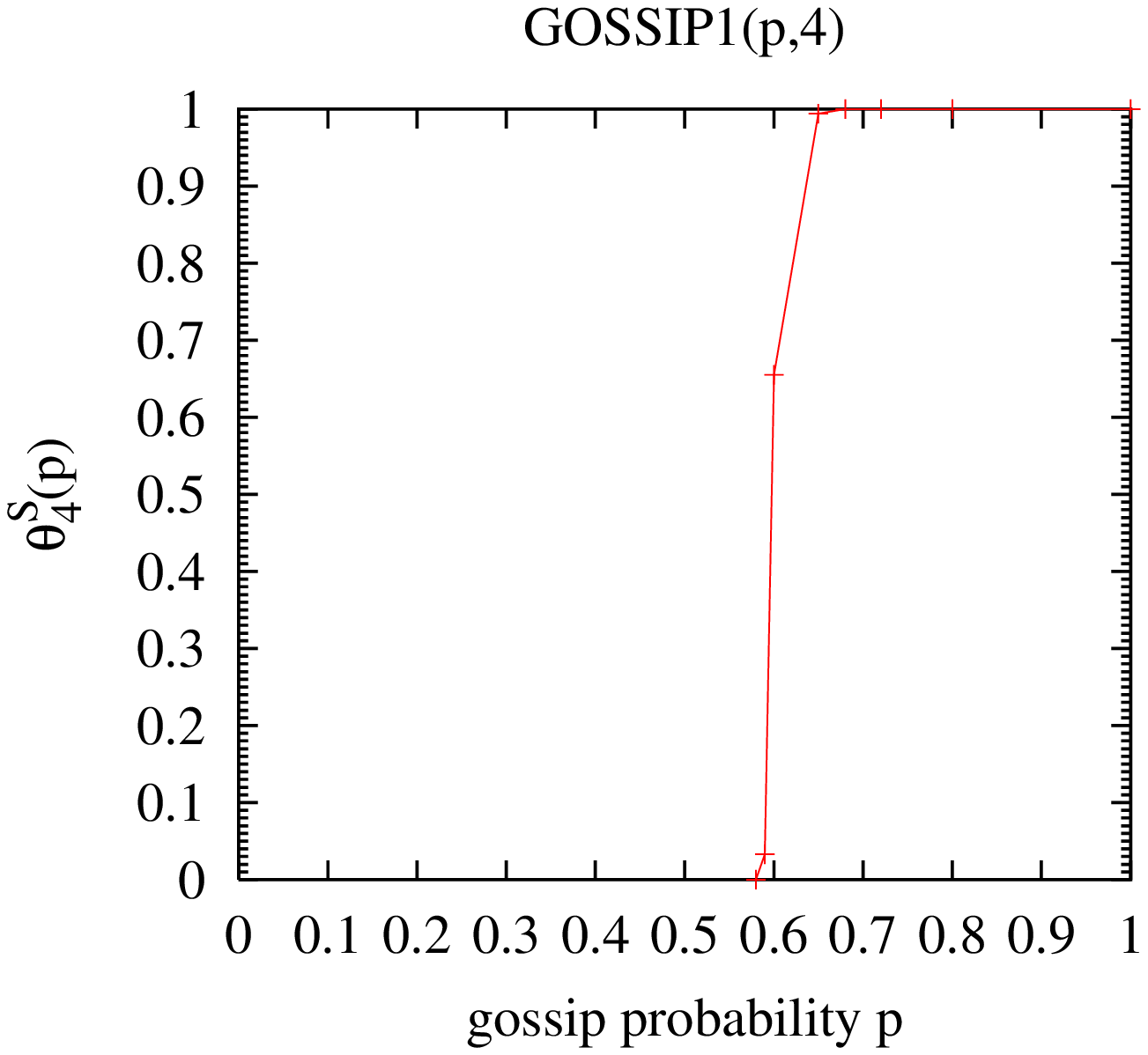} 
\end{tabular}
\end{center}
\caption{$\theta^S_4(p)$ 
as a function of the 
gossip probability $p$ on a 1000x1000 grid.
\label{fig-thetap}
}
\end{figure}

We did one last set of experiments to better evaluate $\theta_k(p)$.
In these experiments, we used 1,000,000  nodes on a $1000 \times 1000$
grid, and placed the source at the center of row 10.  This is far enough
away from the boundary to avoid significant boundary effects.%
\footnote{
%
Experimental results show
that there are nontrivial boundary effects for values of $p$ very close
to $.59$ no matter where we place the source.
Intuitively, this is because for $p$ very close to, but above $.59$,
the probability of having a large disconnecting set of nodes is
nontrivial, and the boundary can help in forming such sets. 
}
The results of using GOSSIP1($p,k$) 
for particular values of $p$ are
illustrated in Figure~\ref{fig-1mgrid}.  As these results show, the 
bimodal effect is very marked by the time we get to such a large
network, and begins to closely approximate the results expected from the
theorem. 
Figure~\ref{fig-thetap} shows how $\theta^S_4(p)$ 
varies with $p$.  As we can see, if $p$ is below $.59$, then the gossip
dies out in almost all executions.  $\theta^S_4(p)$ then increases
very rapidly, going from 0 at .59 to almost 1 at .65.  (The rapid increase
in the case of infinite graphs follows from a deeper mathematical
analysis, and has been discussed in the percolation theory
literature\cite{Grimmett89,meester96}.)  
\input{epsf}
\input{epsf}

Finally, we  considered how $\theta^S_k(p)$ and $\theta^R_k(p)$ varied with
$k$ for a fixed value of $p$.  As theory predicts, $\theta^R_k(p)$ does
not change at all with $p$.  There is some effect on $\theta^S_k(p)$.
Of course, since $\theta^S_1(p) = \theta^S_0(p)/p$, there is a
significant jump as $k$ goes from 0 to 1.  As $k$ increases beyond 1,
there is an increase in $\theta^S_k(p)$, but it is not so significant.
For example, $\theta^S_1(.65) = .95$, $\theta^S_2(.65) = .98$, and
$\theta^S_5(.65) = 1$; similarly, $\theta^S_1(.6) = .53$,
$\theta^S_4(.5) = .67$, and $\theta^S_{10} = .73$.

\section{Heuristics to Improve the Performance of Gossiping}
\label{sec-gossip}

The results of the previous section suggest an obvious way that
gossiping can be applied in ad hoc routing.  Rather than flooding, we
use GOSSIP1($p$,$k$) with $p$ 
sufficiently high to guarantee that almost all nodes will receive the
message in almost all executions.
We can
practically guarantee that the destination node receives the message,
while saving a fraction $1-p$ of messages.  In cases of interest,
where the threshold probability seems to be about  
.65--.75, this means we
can ensure that all nodes get the message using 25--35\% fewer messages
than flooding.
Notice that, if the network is congested and every node has a congestion
dropping probability $f$, then 
to obtain the same results, the broadcast probability needs to be min($p/(1-f)$,1). 
If congestion is very localized, 
then we can simply use $p$ because it
is not likely to change the outcome of a given run of gossiping.
However, the general interaction between gossiping and congestion is a
topic that deserves further study.

\commentout{ 
An obvious question is how to find 
an appropriate 
gossip probability for different networks in practice.
In the networks considered in
the previous section,
we found the appropriate 
probability
by experimentation.
In real applications, we may not have the luxury to perform
such experiments offline.  However, we can achieve the same effect
dynamically.  
Each node initially uses some probability $p_{init}$ as its gossiping
probability.  
The gossiping probability is then adjusted according to the
success/failure of route requests; it
is increased if the route request failure probability is
high, and decreased if the route request failure probability is close to 0. 
To propagate the appropriate probability throughout the network, it  can
be put into the route request packet. Each intermediate node receiving
the packet will gossip with the probability carried in the route
request packet.
Our preliminary experiments have shown that this approach does
produce good behavior, although we do not have enough experience yet
to determine
the best way of making these
adjustment to the gossip probability; we leave this for future work.  
Note that, using such a scheme, it may well be the case that different
nodes in a network use different gossip probabilities.
} 

The basic gossiping scheme can be optimized in a number of ways, using
ideas that have been applied to flooding and ideas specific to
gossiping.  We discuss some optimizations in the remainder of this
section.
This section is intended 
as
a proof of concept
showing that gossiping is a worthwhile approach to explore.
We do not attempt to
do an exhaustive analysis to find the optimal parameters.

\subsection{A two-threshold scheme}

\commentout{
In may cases of interest, a gossip protocol is run in conjunction with
other protocols that give a node a fairly accurate information regarding
its neighbors.
Such information can be obtained, for example, 
by having each node broadcast a
HELLO beacon
to its neighbors from time to time.%
\footnote{The use of the 
HELLO beacon mechanism can be a significant source of overhead, 
so the optimization proposed should be used only in cases where
the necessary information is available or can be obtained with low
overhead.}  
%
If this information is available, 
we can improve the performance of GOSSIP1
further by a simple optimization.  
} 
In many cases of interest, a gossip protocol is run in conjunction
with other protocols. 
If the other protocols maintain fairly accurate
information regarding a node's neighbors, 
we can make use of this information to improve the performance of
GOSSIP1 further by a simple optimization.   

In a random network, unlike the grid, a node may have
very few neighbors.  In this case, the probability that none of the
node's neighbors will propagate the gossip is high.  
In general, we may want the gossip probability at a node to be a
function of its degree, where nodes with lower degree gossip with higher
probability.    
To show the effect of this, we consider a
special case here: 
a protocol with
four parameters, $p_1$, $k$, $p_2$, and $n$.  As in GOSSIP1, 
$p_1$ is the typical
gossip probability and $k$ is the number of hops with
which we start gossiping with probability 1.  The new features are
$p_2$ and $n$; the idea is that 
the neighbors of a node with fewer than $n$ neighbors gossip with
probability $p_2 > p_1$.  That is, if a node has fewer than $n$
neighbors, it instructs its immediate neighbors to broadcast with
probability $p_2$ rather than $p_1$.  Call this modified protocol
GOSSIP2($p_1, k, p_2, n$).
To understand why the {\em neighbors'\/} gossip probability is increased
if there are few neighbors, consider the initiator of the gossip.
Clearly, if none of its neighbors gossip, then the gossip will die.  
If the initiator has many neighbors, even if each gossips with
relatively low probability, the probability that at least one of them
will gossip is high.  This is not the case if it has few neighbors.

\begin{figure}[htb]
\setlength\tabcolsep{0.1pt}
\begin{center}
\begin{tabular}{cc}
\epsfysize=4.0cm \epsffile{mobiFigures/hopdistrndnetn1000p75.ps}  &
\epsfysize=4.0cm \epsffile{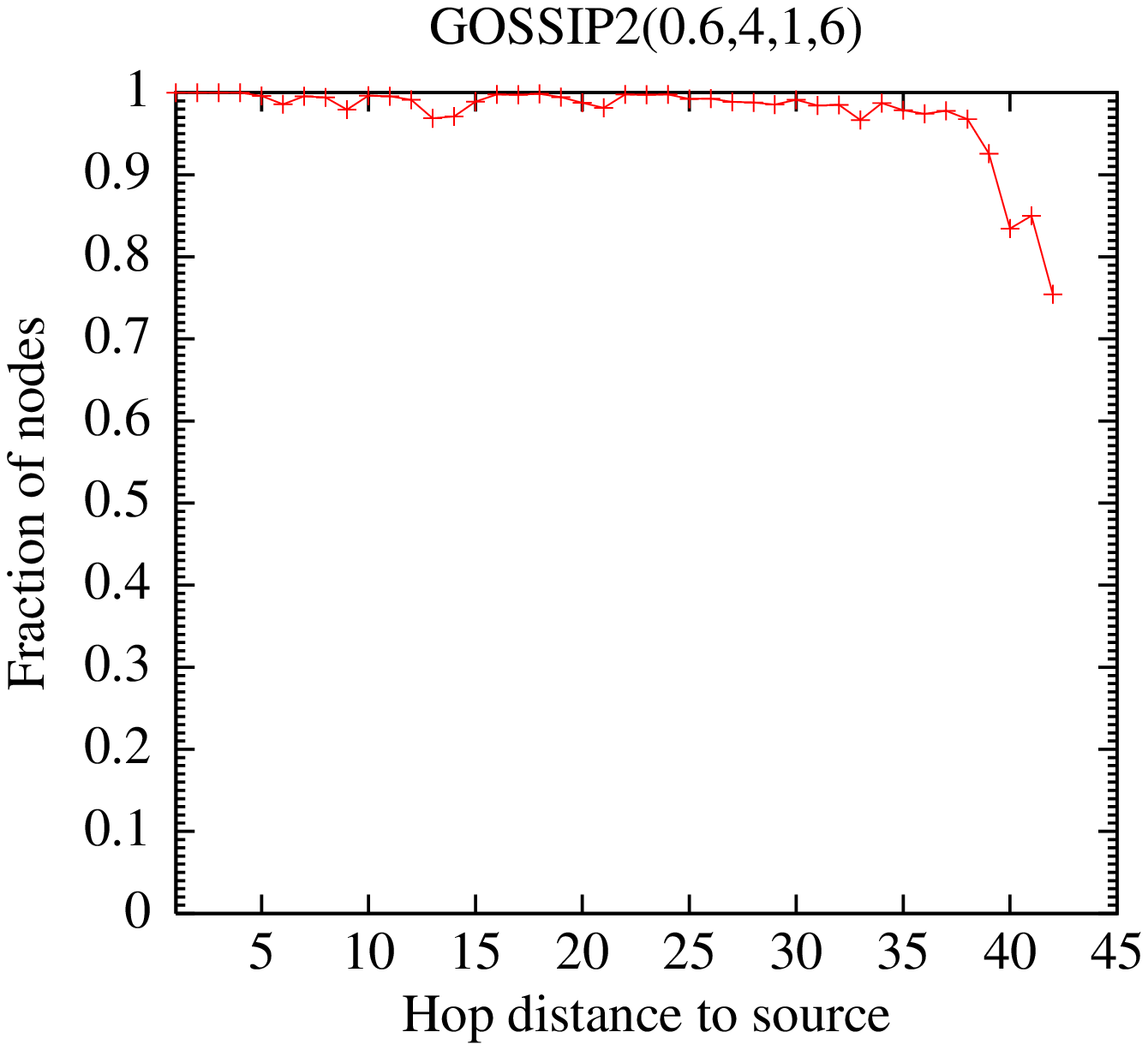} \\
{\footnotesize (a)  } &
{\footnotesize (b)  } \\
\end{tabular}
\end{center}
\caption{Gossiping with two thresholds vs.~one on a random network of
average degree 8.
\label{fig-2threshold}
}
\end{figure}

GOSSIP2 is not of interest in regular networks.  However, in random
networks which typically have some sparse regions, it can have a
significant impact.  
For example, for the random network with average degree 8 first
considered in Figure~\ref{fig-random8}, 
GOSSIP2(0.6,4,1,6) has better performance than GOSSIP1(0.75,4), as shown in 
Figure~\ref{fig-2threshold}, while using 
4\% less messages than GOSSIP1(0.75,4).
Only when $p \ge 0.8$ does GOSSIP1($p,4$) begin to have the same 
performance as GOSSIP2(0.6,4,1,6); however, GOSSIP1(0.8,4) uses  13\% more
messages than GOSSIP2(0.6,4,1,6).

\input{epsf}

There may be other combinations of parameters for GOSSIP2 that give
even better performance; we have not checked 
exhaustively.
The key point is that using a higher threshold
for successors of nodes with low degree seems to significantly improve
performance.

\subsection{Preventing premature gossip death}
As we have seen, the real problem with gossiping is that, if we gossip
with too low a probability, the message may ``die out'' in a certain
fraction of the executions.  
Measures can be taken to
prevent this (for example, having successors of nodes with low degree
gossip with a higher probability) but, unfortunately, there is no way
for a node to know if a message is dying out.  Nevertheless, a node may
get some clues.  One such clue is not getting too many copies of the
message.   Suppose that a node $x$ got the message but does not
broadcast it because its coin toss landed ``tails''.  Further
suppose that $x$ has $n$ neighbors.  If the message does not die out,
then it would expect that all of its neighbors would get the message as
well, and thus, if the gossip probability is $p$, it should get roughly
$pn$ messages from its neighbors.  If it gets significantly fewer than
$pn$ within a reasonable time interval, then this is a clue that the
message is dying out.   

This suggests the following optimization of GOSSIP1 and GOSSIP2.
If a node with $n$ neighbors receives a message and does not broadcast
it, but then does 
not receive the message from at least $m$ neighbors within a
reasonable timeout period, it broadcasts the message to all its
neighbors.  The obvious question here is what $m$ should be.
If $m$ is chosen too large, then we may end up with too many messages.
Our experiments show that we actually get the most significant
performance improvement by taking $m=1$.  Let GOSSIP3($p,k,m)$ be just
like GOSSIP1($p,k$), except for the following modification.
A node that originally did not broadcast a received message (because its 
coin landed tails), but then did not get the message from at least $m$ 
other nodes within some timeout period, 
broadcasts the message immediately after the timeout period.
(The choice of timeout period can be taken quite small. We discuss this
issue in details in Section~\ref{sec-aodv}.)
It may seem that such rebroadcasting can significantly effect the
latency of the message.  However, as the experiments discussed below
show, if the parameters are chosen correctly, 
latency is not a problem at all. 


\input{epsf}
\begin{figure}[ht]
\begin{center}
\begin{tabular}{c}
\epsfysize=4.0cm 
\epsffile{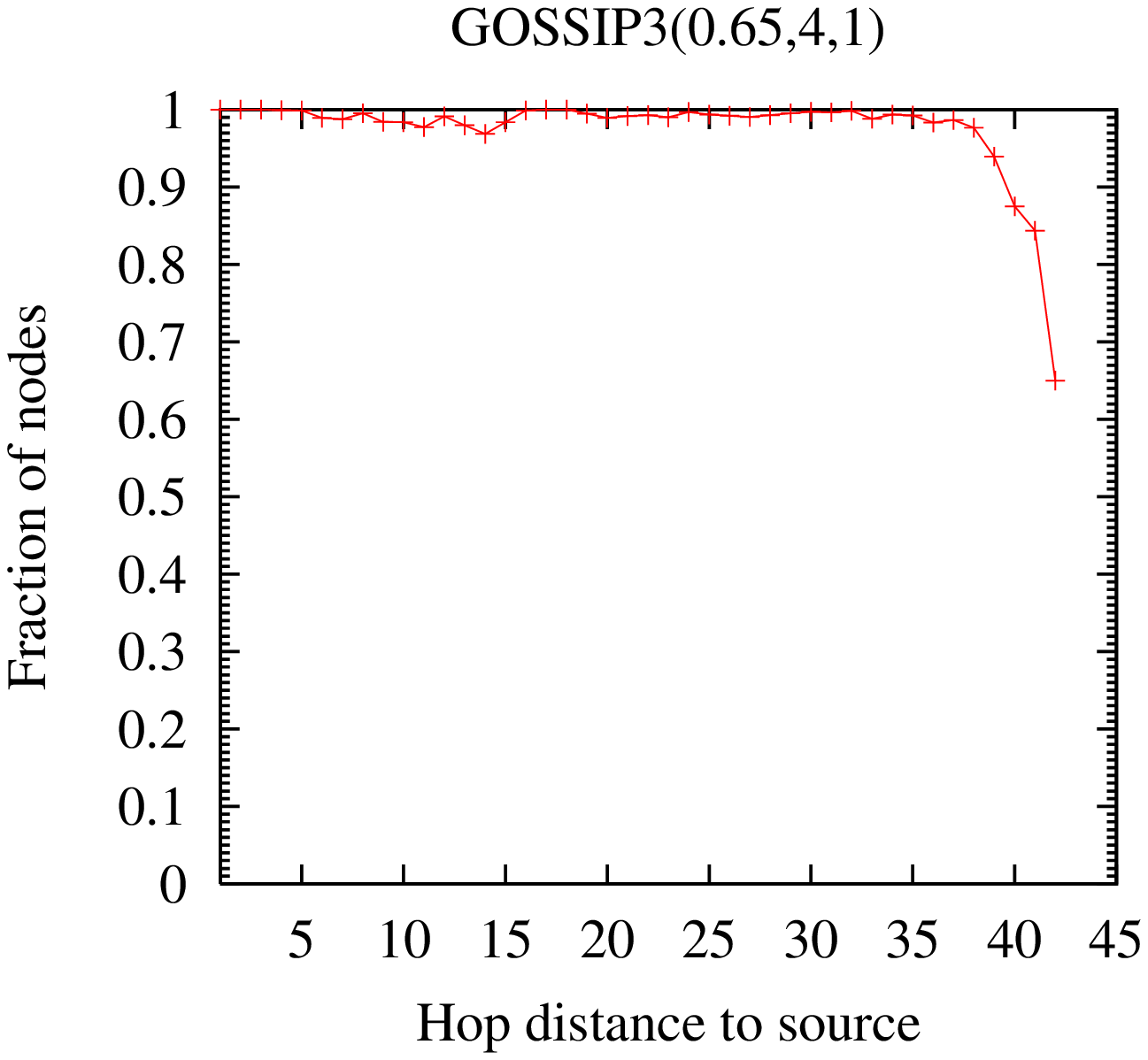} 
\end{tabular}
\end{center}
\caption{GOSSIP3 on a random network of degree 8.
\label{fig-PPGD}
}
\end{figure}
As Figure~\ref{fig-PPGD} shows,
the performance of 
GOSSIP3(0.65,4,1) is even better than that of GOSSIP1(0.75,4).
However, GOSSIP3(0.65,4,1) sends only  67\% of the messages sent by
flooding.  By way of contrast, 
GOSSIP1(0.75,4)
sends 75\% of the messages sent by flooding. 
Thus, we get better performance using GOSSIP3 while sending 8\% fewer
messages.  

To examine the effect of GOSSIP3 on latency, we 
recorded the number of timeout intervals a message
experienced, using a variable $L$, which was augmented every time a
message was forwarded after a timeout.
Among all the messages sent by
GOSSIP3(0.65,4,1), only 2\% have $L \ge 1$. Among these messages with
$L \ge 1$, 95\% of them have $L \leq 2$.   Thus, it seems latency is not
a problem here.



\subsection{Retries}
The bimodal distribution observed in the use of gossiping can be
viewed as a significant advantage.  Once a route is found,
acknowledgments are propagated back to the source along the route, so
the source knows the route.  If a route is not found within a certain
timeout period, there are two possibilities: either there is no route
at all, or the protocol did not detect it.  Our focus is on networks
that are sufficiently well connected that there typically is a route.
However,
when using a gossiping protocol, there is always a
possibility that a route will not be found even if it exists.
Of course, there is a
simple solution to this problem: simply retry the protocol.  
Thus,
for example, the probability of finding a route 
within two attempts
to a node at distance 25 using GOSSIP1(.65,4) in the random network
with average outdegree 8 is .95:
the probability of a node not receiving a message in any given
execution 
of the protocol
is .23, and executions are independent.

With retries,
the bimodal message distribution works
significantly to our benefit.  As we observed, with GOSSIP1(.65,4), in
72\% of the executions, almost all nodes get the message.  If we pick a
destination at random, in those executions where almost all nodes get
the message, the destination 
is likely to 
get the message and a retry will not
be necessary.  On the other hand, in those executions where hardly any
nodes got the message, a retry will probably be necessary.
However, such failing
gossip attempts do not involve too many transmissions, since
most nodes do not get the message 
in any case.

Of course, retries increase latency, even if they do not significantly
increase the number of messages sent.
This is especially true in large networks,
where the timeout period will have to be large so as to allow the
message to propagate throughout the network.
However, even here, 
the bimodal distribution can be used to 
advantage to decrease the
retry latency.
Note that each message must keep 
track of 
the number of hops 
it has taken.  We can modify the algorithm so as to
require that any node that receives a message with, say, 15 hop counts,
forwards an acknowledgment to the sender along that route with some
probability.  (The probability can be chosen so that the sender receives
an expected number of, say, five acknowledgments if almost all nodes
get the message.)  Because of the bimodal distribution, if the sender
receives several acknowledgments, then it can be fairly confident
that the execution is one in which almost all nodes are getting the
message.  On the other hand, if it does not receive several
acknowledgments, it is likely that the execution is one in which hardly
any nodes get the message, and it should resend the message
immediately.  This shows that we can bound the latency of retry,
independent of the network size. 

%

\commentout{
\subsection{Temporal, spatial, and topological optimizations}
Of course, gossiping can also be combined with all the structural and
locality optimizations discussed in the introduction. For example, the
topological locality optimization in \cite{Das99} can be easily
incorporated into gossiping. 
If the source can estimate the $d_{max}$ parameter mentioned in
Section~\ref{sec-introduction},  gossiping can stop after 
$d_{max}+k$ hops, where $k$ is a small constant that takes care of
boundary and back-propagation effects.
be applied to gossiping straightforwardly.
if an intermediate node has a route that is no older than some time
duration  before the current route request, the node can choose to
unicast the route request to the route request source instead of
continuing the gossiping process.

}

\subsection{Zones}\label{zones}
One of the best-known 
optimizations to flooding is the 
{\em zone routing protocol\/} (ZRP)~\cite{ZRP98}.
In ZRP, each node $u$ maintains a so-called {\em zone}, which
consists of all the nodes that are at most $\rho$ hops away from
$u$, for some appropriately chosen {\em zone radius\/} $\rho$.
A node that is exactly $\rho$ hops away from $u$ is called a {\em
peripheral node\/} of $u$.

A node proactively tries to maintain complete routing tables for all
nodes in its zone.  
Initially, a node discovers who its neighbors
are and then broadcasts its neighbors to its zone (by using flooding up to
hop count $\rho$).  Then each time it discovers a change (i.e., that it
has lost or gained a neighbor), it broadcasts an update.
This procedure ensures that a node has a very accurate picture of its zone.

If a source wants to send to a destination in its zone, it simply routes
the message directly there, since it knows the route.  
Otherwise, it sends a route request query to the peripheral nodes in its
zone. If the destination is in a peripheral node's zone, the peripheral node 
replies with the route to the query originator. Otherwise, it forwards the query to
its peripheral nodes, 
which in turn forwards it, and so on.

\commentout{
The route to a destination that is outside a node's
zone is found on-demand 

using a protocol called the {\it
Interzone Routing Protocol (IERP)}. Rather than using flooding to
``blindly'' forward route queries from a node to its neighbors, IERP 
{\em bordercasts\/} 
route queries to a node's peripheral nodes. Bordercast is done through
the construction of a bordercast tree $T$ rooted at the bordercast
source. In order for all the nodes on the bordercast tree to have the
same view, the bordercast source puts the tree $T$ in the route
query packet.  
This may have the disadvantage, however, of increasing the
message size considerably.  Alternatively, rather than sending to
peripheral nodes, the bordercast source can bordercast the message to
nodes distance $\lfloor \rho/2 \rfloor$ away.  Since all the nodes on
the bordercast tree have a common view of the topology of the network in
a zone of radius $\lfloor \rho/2 \rfloor$ around the bordercast source,
there is no need to include the tree in the message in this case.
However, this approach loses some of the benefits of the zone.
In any case, when a leaf in the bordercast tree gets the message, it
repeats the process: if the destination is in its zone, it send the
message directly to the destination; otherwise, it bordercasts.
}

In the context of ZRP, there are two advantages of maintaining a zone.
First, if a node is in the zone, 
flooding is unnecessary; a message can be sent directly to the intended
recipient, saving much control traffic. 
This
brings about a significant improvement in overall performance if a
substantial fraction of nodes are in the zone
(which is likely to be true in a small network, but far less likely in a
large one).
Second, if we want to
send a message outside the zone, we can multicast to the boundary of
the zone (or a subset of the nodes on the boundary), which can be a
significant saving over flooding. 
However, there is a tradeoff in choosing the size of the zone: a
bigger zone benefits more from these two advantages, but also results in
overhead for proactive maintenance of the zones. 
In general, the optimal zone size will depend on factors like mobility
and 
frequency of route requests.

The idea of zones can be used in gossiping as well.  Here there is a
third advantage: if a node in the zone receives a gossip message, then
it can send it directly to 
any node in the zone.
This means that it would suffice for a gossiping protocol to get the
message to a node in the intended recipient's zone.
How much of an advantage is this?  
In large networks, the advantage is quite minimal.
As we have observed, 
gossiping is essentially bimodal: 
for typical gossip probabilities, either hardly any nodes get the
message or most of them do.
Zones have a relatively
small effect in either case.  Thus, zones help only in the relatively few
executions that exhibit ``intermediate'' behavior.  
\input{epsf}
\begin{figure}[htb]
\setlength\tabcolsep{0.1pt}
\begin{center}
\begin{tabular}{cc}
\epsfysize=4.0cm \epsffile{mobiFigures/hopdistrndnetn1000p65.ps}  &
\epsfysize=4.0cm \epsffile{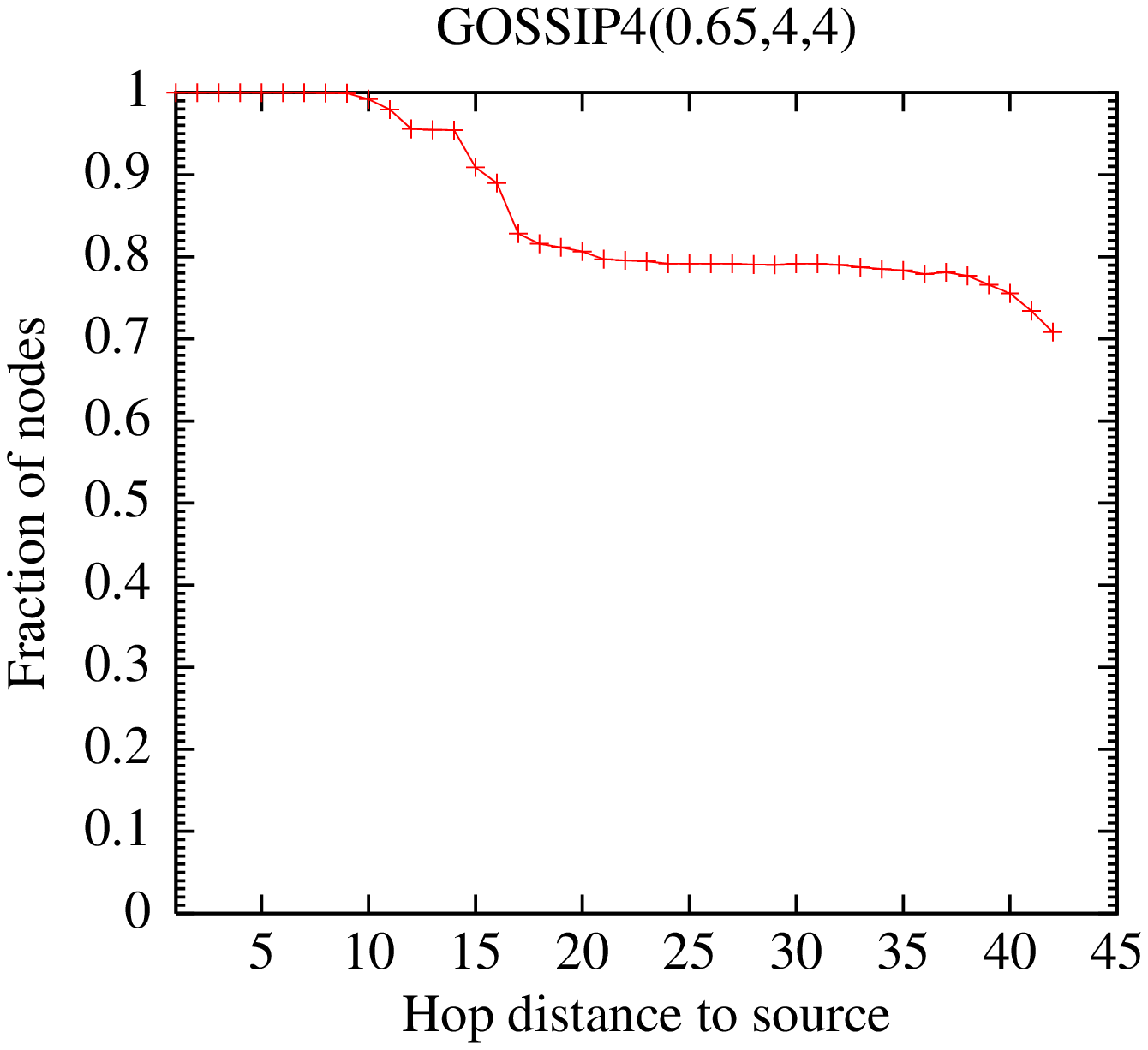} \\
{\footnotesize (a)   } &
{\footnotesize (b)   } \\
\end{tabular}
\begin{tabular}{c}
\epsfysize=4.0cm \epsffile{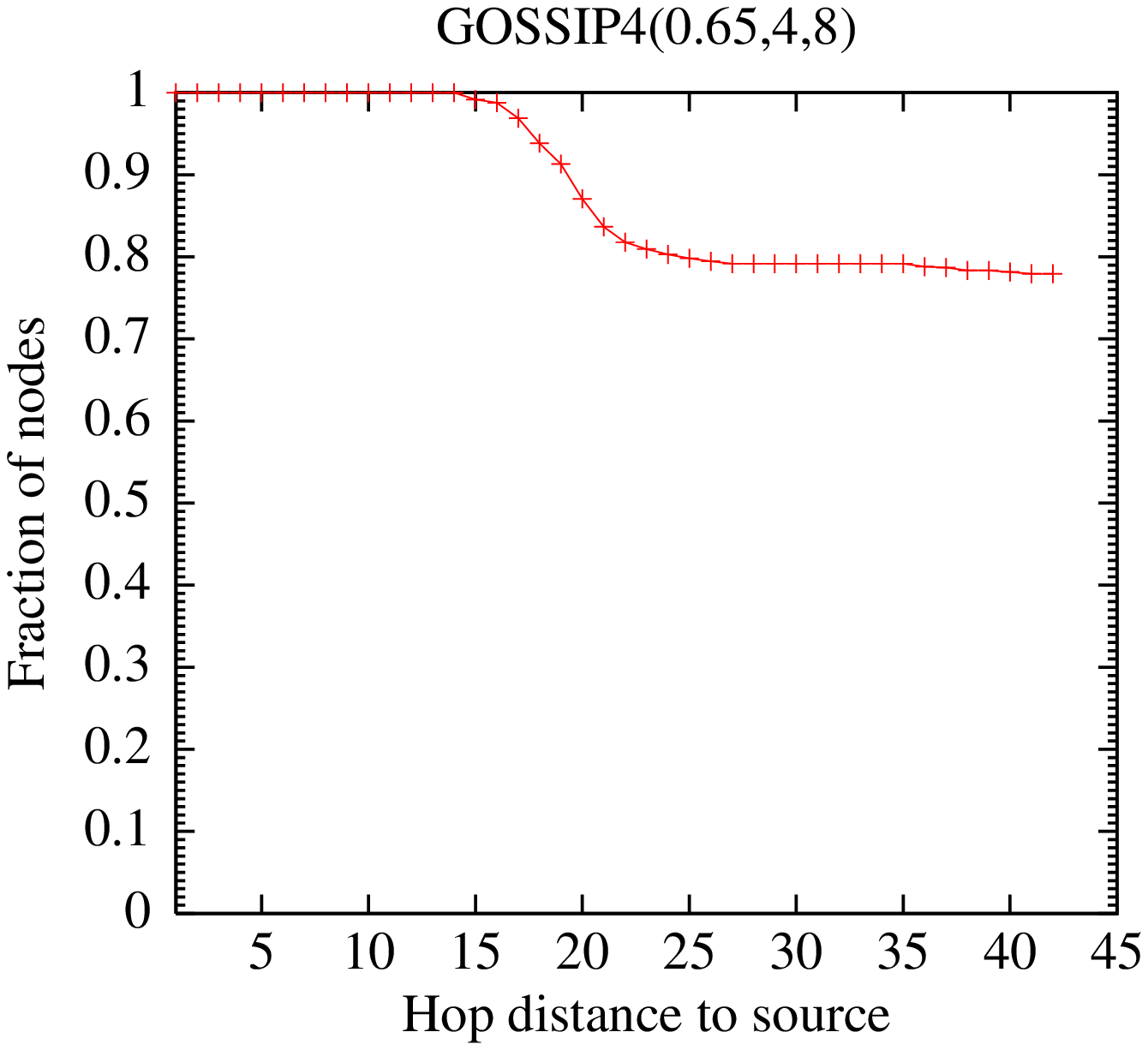}  \\
{\footnotesize (c)   } 
\end{tabular}
\end{center}
\caption{Gossiping with zones on a random network of average degree 8.
\label{fig-zone}
}
\end{figure}
Let GOSSIP4($p,k,k'$) be just like GOSSIP1($p,k$),
except that each node has a zone of radius $k'$.  
Comparing Figure~\ref{fig-zone}(b) to
Figure~\ref{fig-zone}(a), we see using a zone radius of 4 
with gossiping probability .65 in the random network with average
degree 8
improves
performance by only a few percent over most of 
the distances.
However, it
does ameliorate the back-propagation effect.  As shown in
Figure~\ref{fig-zone}(c), increasing the zone radius to 8 does not
significantly improve the limiting performance, but it has an even more
beneficial effect on the back-propagation problem.

\input{epsf}
\begin{figure}[htb]
\setlength\tabcolsep{0.1pt}
\begin{center}
\begin{tabular}{cc}
\epsfysize=4.0cm \epsffile{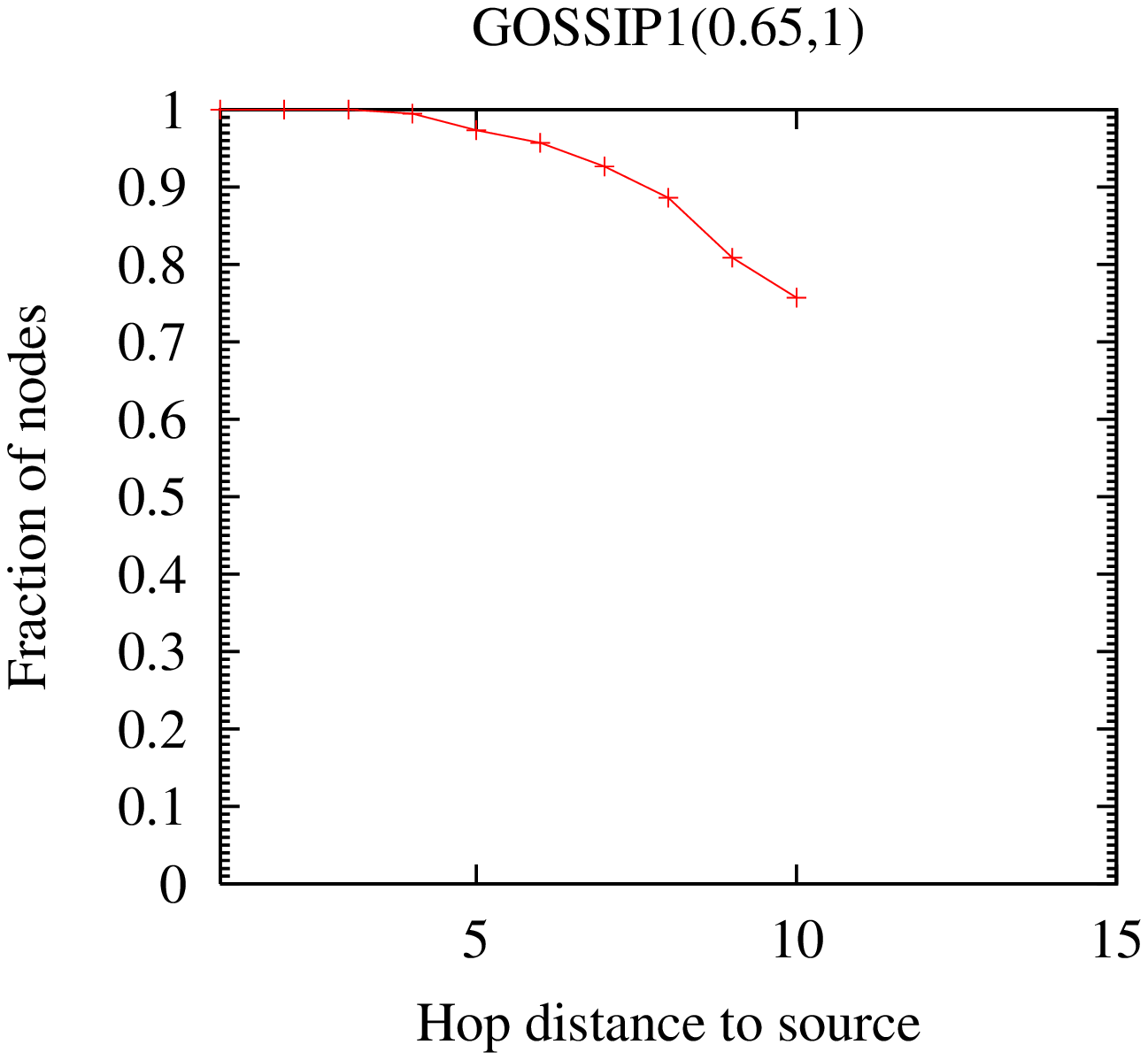} &
\epsfysize=4.0cm \epsffile{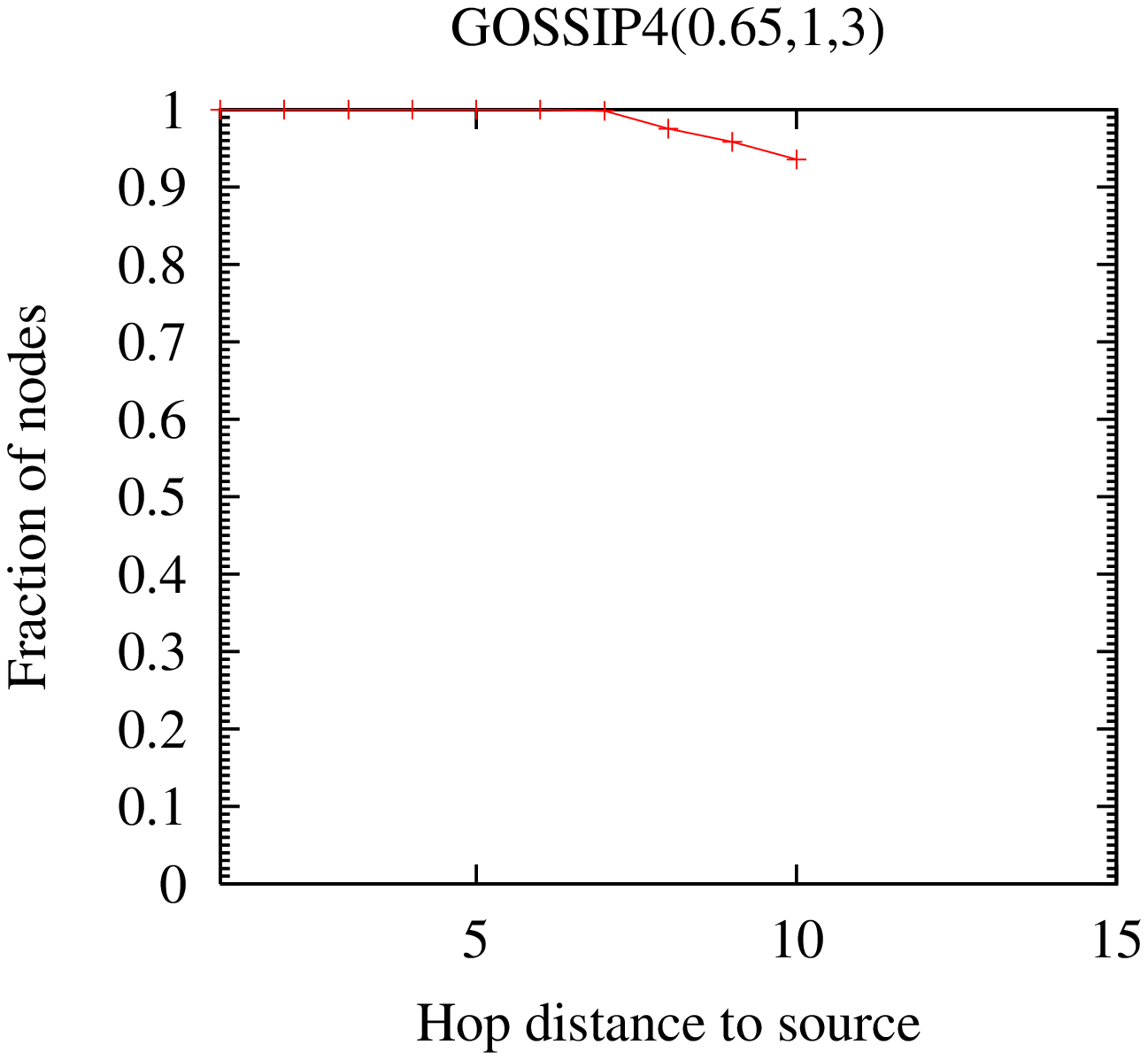} \\
{\footnotesize (a)   } &
{\footnotesize (b)   } 
\end{tabular}
\end{center}
\caption{Gossiping with zones on a 100 node random network.
\label{fig-zone100nds}
}
\end{figure}

The situation is much different for smaller  networks.
Here zones can have a significant impact.
For example, if we use gossip probability .65 in a random network with
100 nodes and average degree 13, 
the network is too small 
for the bimodal effect to show up.  However, the back-propagation problem
is significant. 
As Figure~\ref{fig-zone100nds} shows,
for the small random network of 100 nodes, if we use GOSSIP1(0.65,1),
then only 76\% of nodes at distance 10 get the message. However, if
we have a zone of radius 3 (GOSSIP4(0.65,1,3)), then 96\% of nodes at
distance 10 get the message. 



\section{Incorporating Gossiping in AODV}
\label{sec-aodv}
How much does gossiping really help in practice?  That depends, of
course, on issues like the network topology, mobility, and how
frequently messages are generated.  We believe that in larger networks
with high mobility 
many
of the optimizations discussed in the
literature will be much less effective.  
(We discuss this point in more
detail below in the context of AODV.)  In this case, flooding will occur
more frequently, so gossiping will be particularly advantageous.
However, as our results show, gossiping can provide significant
advantages even in small networks. 

To test the impact of gossiping, we considered AODV, 
one of 
the
best-studied ad hoc routing protocols in the literature.
We compared pure AODV to a variant of AODV that uses gossiping instead
of flooding whenever AODV would use flooding.  
We do not have the resources to simulate 
the protocols in large
networks.  However, our results do verify the intuition that, with high
mobility (when flooding will be needed more often in pure AODV),
gossiping can provide a significant advantage.

\subsection{A brief overview of AODV}
\label{sec-AODVoverview}
Using AODV,
the first time a node $u$ requests a route to node $v$, 
it
uses an {\em
expanding-ring search\/} to find the route.  That is, it first tries to
find the route in a zone of small radius, by flooding. It then tries to
find the route in zones of larger and larger radius.  If all these
attempts fail, it resorts to flooding
the message through the whole network.
The exact choice of zone radii
to try is a parameter of AODV.  Typically, not too many radii are
considered before resorting to flooding
throughout the  network.

AODV also maintains a routing table where it stores the route after it
has been found.
If AODV running at node $u$ gets any packet with source $u$ and
destination $v$,
the route in the routing table will be tried first.
If any 
node $w$ on the route from $u$ to $v$ detects that the link to the next
hop is down, then 
$w$ generates a route error (RERR) message,
which is propagated back 
to $u$.
When $u$ receives the RERR message, it deletes the route to
$v$ from its routing table.


\subsection{GOSSIP3 in AODV}
We added gossiping to AODV in a particularly simple way.   If 
the expanding-ring search with a smaller radius fails, rather than
flooding to the whole network, 
we use GOSSIP3(.65,1,1).  
(We used these parameters since they gave good performance in
the particular scenarios we considered.)
The timeout period of GOSSIP3 should be big enough to allow
neighboring nodes to gossip. The $NODE\_TRAVERSAL\_TIME$ parameter of
AODV is a conservative estimate of the average one hop
traversal time for packets that includes queueing delays,
interrupt processing times and transfer times. In our experiments, we
set the timeout interval to be $i*NODE\_TRAVERSAL\_TIME$
where $i$ is a small integer ($i=5$ in our reported results). 
Note that we do not use
GOSSIP3 in the expanding-ring search with a smaller radius. Because of the
back-propagation effects,  flooding is actually more efficient than
gossiping for a zone with a small radius.
We call the variant of AODV that uses GOSSIP3 AODV$+$G. 

\subsection{Simulation model and performance results}
Our simulation is done in the ns-2 \cite{ns_2} simulator. This is also
the simulator the literature uses to evaluate AODV. 
We use the AODV implementation in ns-2 downloaded from 
the web site of one of its authors,
using IEEE 802.11 as the MAC layer
protocol. The radio model simulates Lucent's WaveLAN
\cite{lucent-wavelan} with a nominal bit rate of 2Mb/sec and a
nominal range of 250 meters. The radio propagation model is the
two-ray ground model \cite{rap96}.

Our application traffic is CBR (constant bit rate). The
source-destination pairs (connections) are chosen randomly.  The
application packets are all 512 bytes.   
We assumed a sending rate of 2 packets/second and 30 connections.

For mobility, we use the {\em random waypoint} model \cite{joch98}
in a rectangular field. The simulation scenarios are as follows: 150
nodes are randomly placed in a grid of $3300 m \times 600 m$.
We chose this layout because in some sense it provides a worst-case
estimate of the performance of gossiping.  For this layout the gossip
threshold is about .65.  With other more ``square'' layouts, such as
$1650 \times 1200$, it is possible to gossip with lower probability
(closer to .5), so the saving due to gossiping will be even more
significant.  
%
There
are 30 connections, each generating 2 packet/sec; simulation time is
525 seconds; each node moves with a randomly chosen speed (uniformly
chosen from 0-20 m/sec), then pauses for $\tau$ seconds after reaching
a randomly set destination. We vary the pause time to simulate
different mobility
scenarios.
Each data point represents an average of five
runs using the identical traffic model, but with different randomly
generated mobility scenarios. To preserve fairness, identical mobility
and traffic scenarios are used for both AODV and 
AODV$+$G.

We used the same configuration parameters for AODV as those 
used in \cite{DasPerk00}. 
Of particular interest to us are the
expanding-ring search parameters. In the ns-2 implementation of AODV,
first a zone radius of 5 hops is tried; 
if no route is found,
network-wide flooding is used.

We study the performance of the following four metrics, of which the first three
were also studied in \cite{DasPerk00}: 
\begin{itemize}
\item
The {\em packet delivery fraction} represents the ratio of the number of
data packets successfully delivered to the number of data packets 
generated by the CBR sources. 
\item The {\em average end-to-end delay} of data packets
includes all possible delays caused by buffering during routing
discovery, queuing at the interface queue, retransmission at the MAC
layer, propagation, and transfer time. 
\item The {\em normalized
routing load} represents the number of routing packets transmitted per
data packet delivered at the destination. Each hop-wise packet
transmission is counted as one transmission. 
\item The {\em route length ratio\/} compares the shortest route length found
to the actual shortest route length.
\end{itemize}

\input{epsf}
\begin{figure}[htb]
\setlength\tabcolsep{0.1pt}
\begin{center}
\begin{tabular}{cc}
\epsfysize=4.0cm \epsffile{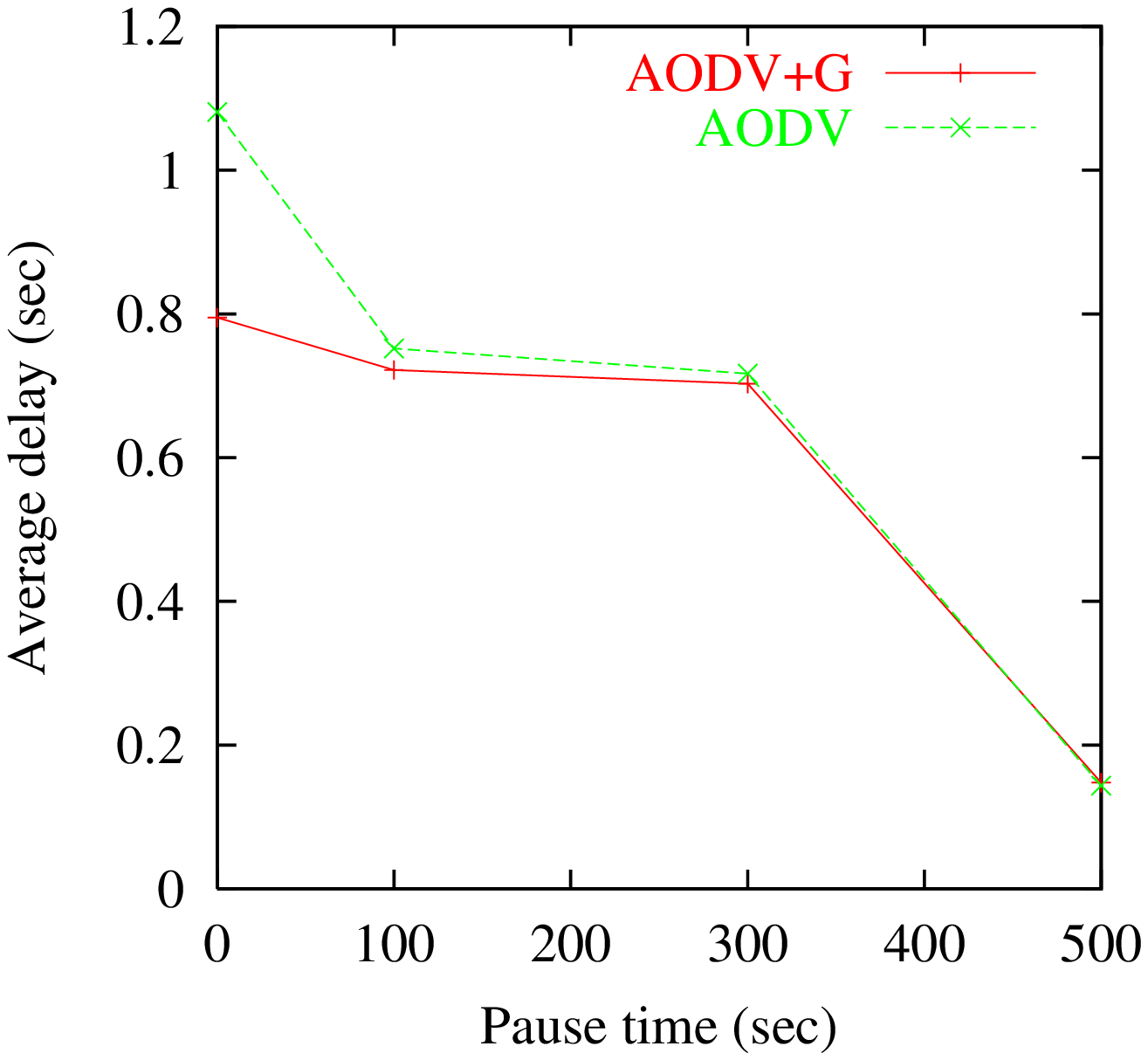} &
\epsfysize=4.0cm \epsffile{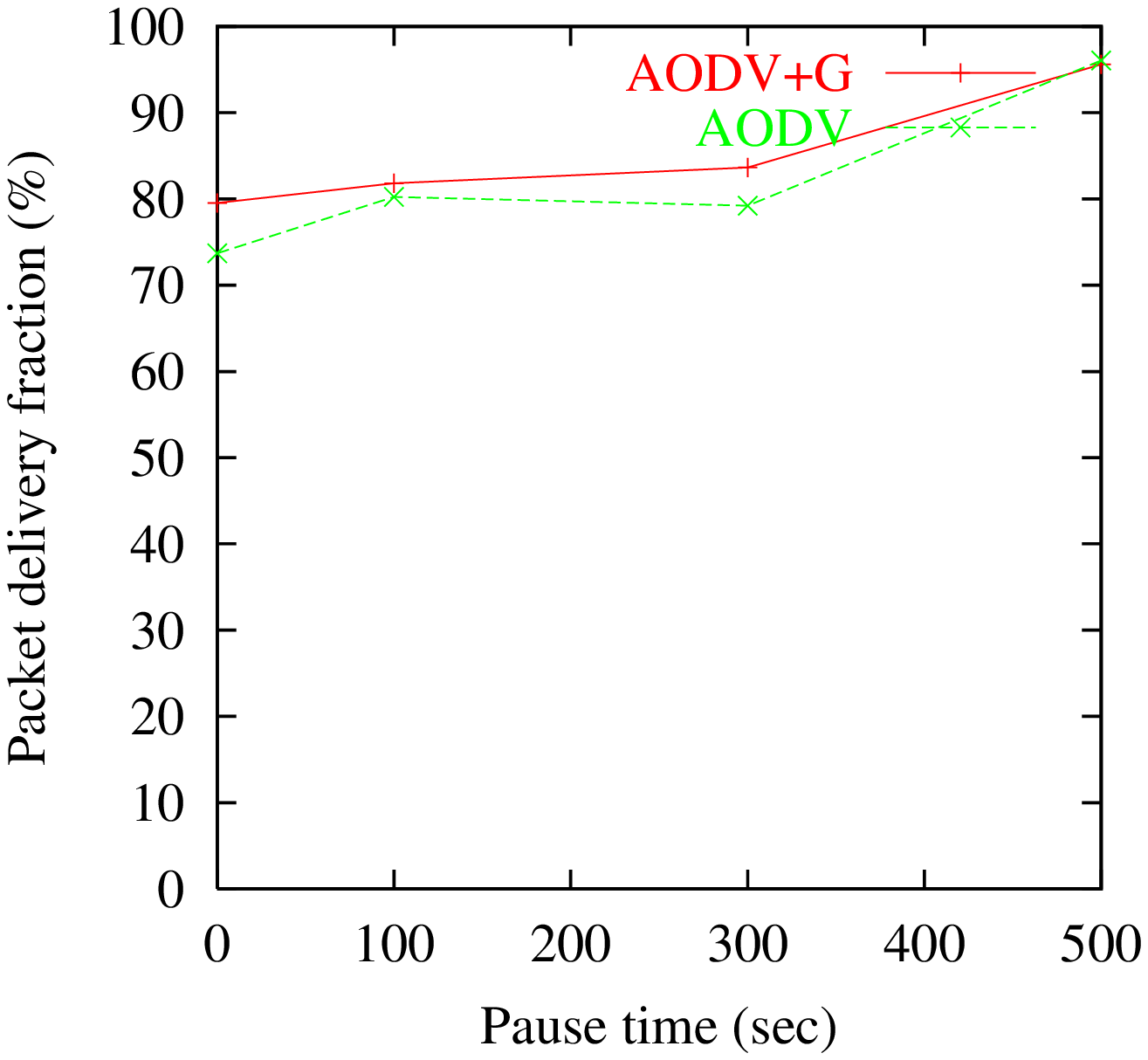} \\
{\footnotesize (a)   } &
{\footnotesize (b)   } \\
\epsfysize=4.0cm \epsffile{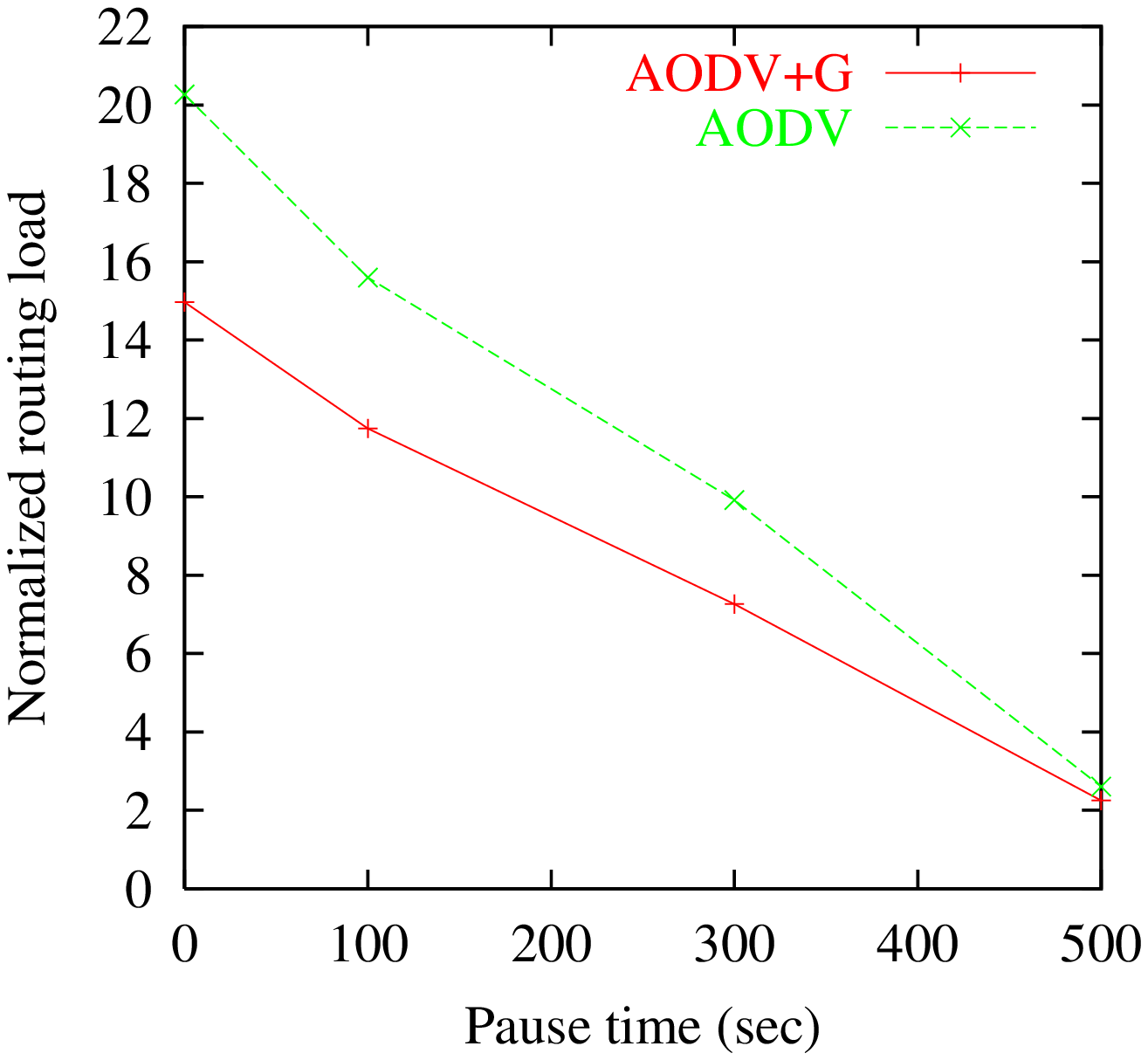} &
\epsfysize=4.0cm \epsffile{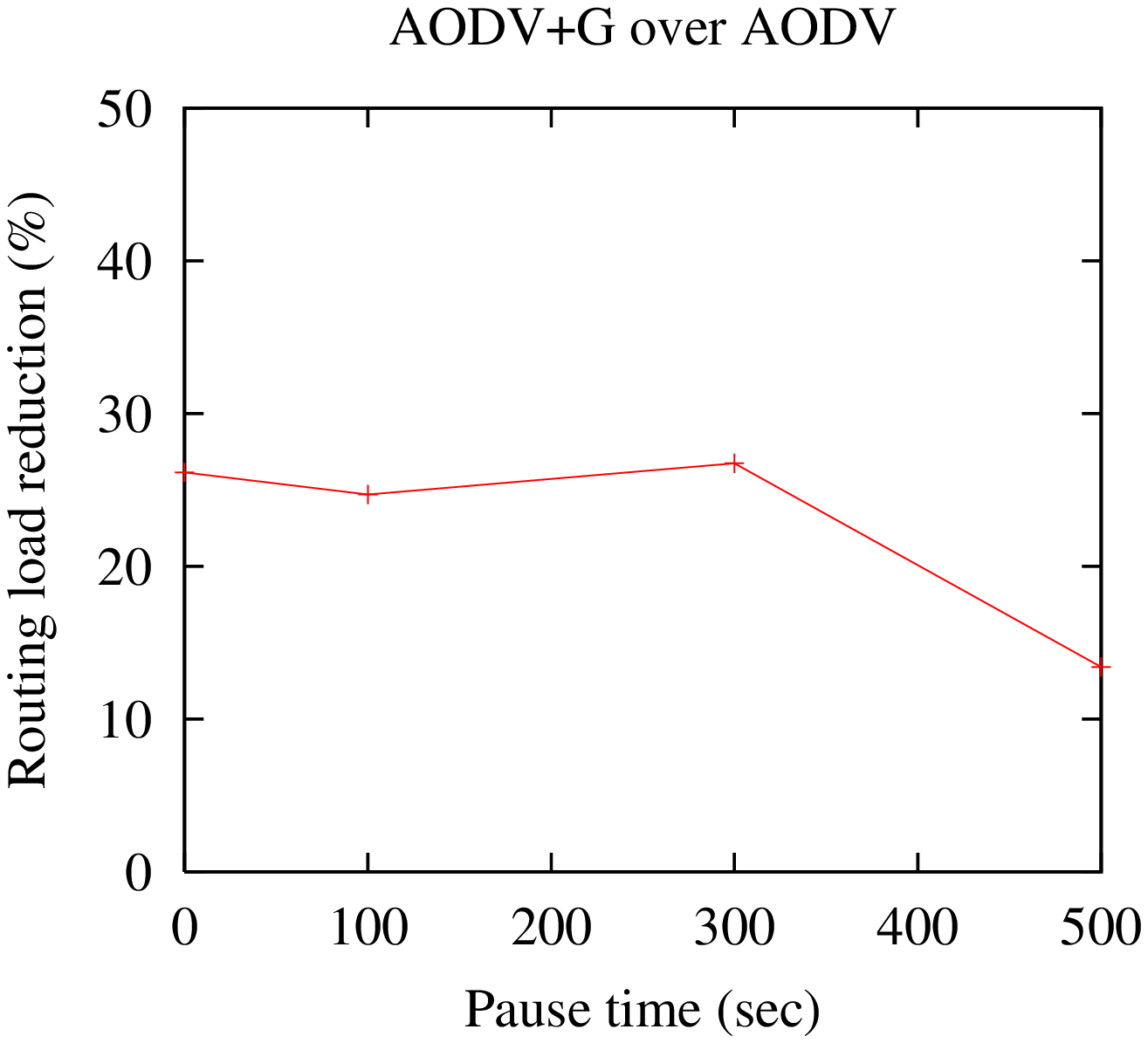} \\
{\footnotesize (c)   } &
{\footnotesize (d)   } \\
\epsfysize=4.0cm \epsffile{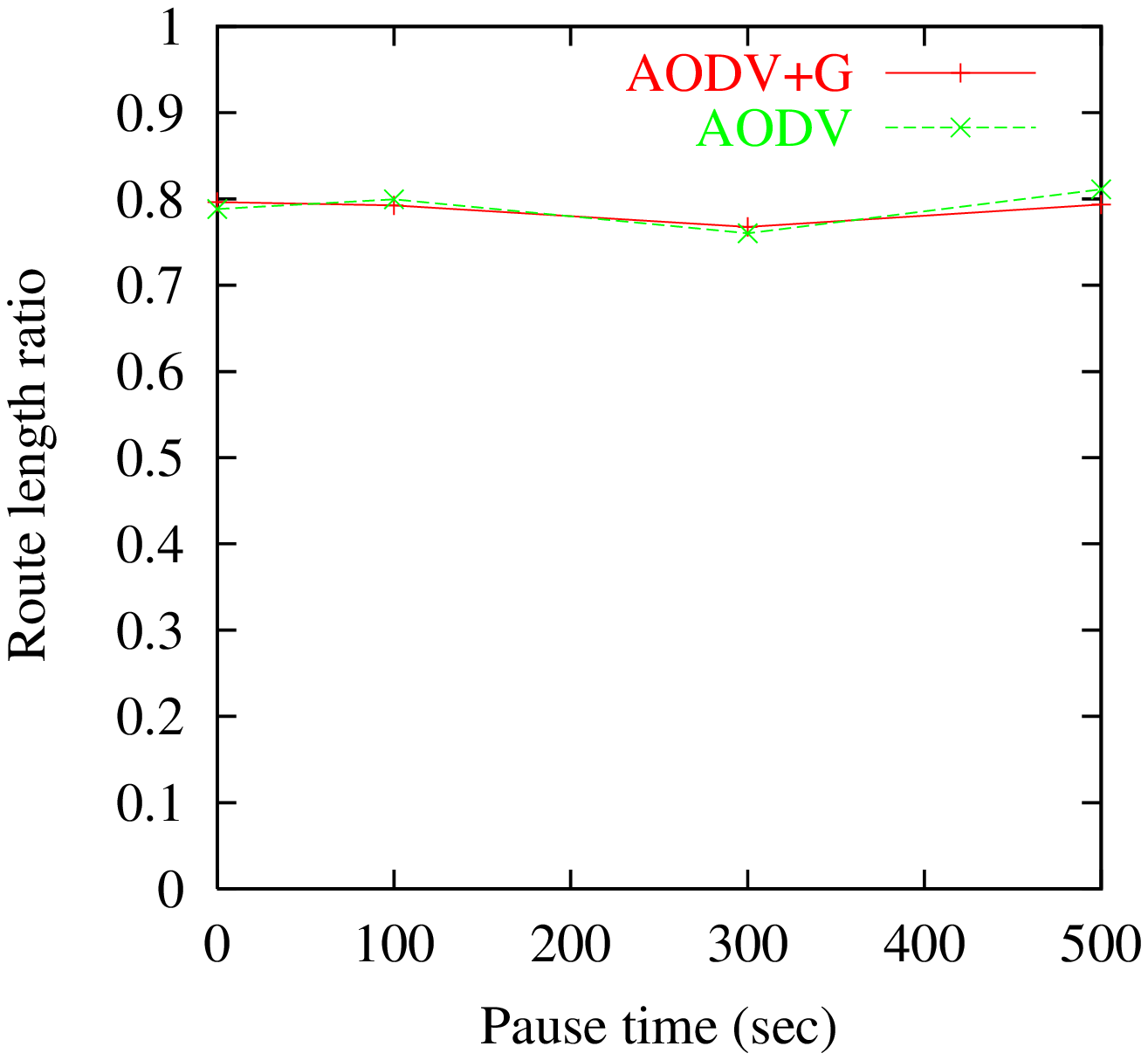} & \\
{\footnotesize (e)   } &
\end{tabular}
\end{center}
\caption{ AODV$+$G vs.~AODV.
\label{fig-aodv}
}
\end{figure}

From Figure~\ref{fig-aodv}(a) and~\ref{fig-aodv}(b), we see that
AODV$+$G delivers better network performance than AODV in terms of
end-to-end delay and packet delivery fraction. The performance
improvements correlate with the amount of routing load reduced. This
is not surprising, since routing load increases with mobility and
constitutes a significant part of the network load (as can be seen from
Figure~\ref{fig-aodv}(c)). 
At pause time 0, AODV$+$G reduces average end-to-end delay by 36\% and
increases throughput by 8\%.  From Figure~\ref{fig-aodv}(c)
and~\ref{fig-aodv}(d), we see that AODV$+$G  
reduces the routing load; the reduction is from 14\% to 27\% in terms
of normalized routing load. 

Finally, we consider route lengths.  Note that neither gossiping nor
flooding 
(as used by AODV)
will necessarily find the shortest route.  For example, suppose
that $(u_0, u_1, u_2,u_3)$ is the shortest path from $u_0$ to $u_3$, but
that there is another path $(u_0, v_1, v_2, u_2, u_3)$.  It is possible
that after $u_0$ broadcasts a route request, $u_2$ will receive it 
along the path from $v_2$ before receiving it from $u_1$.  
Since, in AODV, $u_2$ would save in its routing table information from
only the first route request to arrive, 
AODV will not necessarily discover the
shortest route.
For similar reasons, with gossiping, we may not always discover
the shortest routes.  
Our experimental results show that in the 150-node network studied here,
the length of paths found by flooding and by our gossiping algorithm are
essentially indistinguishable.   We considered
the ratio of the shortest route found by AODV to the
actual shortest route, and similarly for AODV$+$G.  
Figure~\ref{fig-aodv}(e) shows that the routing length ratio for
AODV$+$G and AODV is almost the same (and, indeed, is sometimes
marginally better for AODV$+$G).
However, this result seems to some extent to be an
artifact of the particular small network and the gossip probability
used  here.  Experimental results performed on the networks studied
in Section~\ref{sec-bimodalexp} show that gossiping finds routes 10-15\%
longer than flooding if gossiping is done with a probability just a little
above threshold.   
The gap decreases as the gossiping probability increases; 
for sufficiently large gossip probability, the route lengths are again
essentially indistinguishable.
%

These simulations were carried out in a network with 150 nodes.  
In such a small network, even if route-destination pairs are chosen at
random, a great many pairs will be within 5 hops of each other, and will
thus be discovered by the expanding-ring search.
Indeed, in our simulation, roughly 30\%-40\% of the routes discovered
had a length of less than or equal to 5.
Thus, as many as 40\% of the routes are discovered by
the expanding-ring search.
We expect that things will be quite different in a larger network.
Of course, this depends in part on the nature of route requests and 
the choice of parameters for the expanding-ring search.  While it is
possible that many requests will be local, 
there are applications for which this seems unlikely.
Certainly if route-destination pairs are chosen at random, then
expanding-ring search is unlikely to be effective for almost any choice
of parameter settings. That is, a great many source-destination pairs are likely
to be far apart, so no expanding-ring search is likely to find them
efficiently.  
Additionally, expanding-ring search may add a great deal of
routing traffic and route discovery latency. 
By way of contrast, gossiping continues to perform well in large networks.
Thus, we predict
that the relative advantage of AODV$+$G over pure AODV will increase as the
network gets larger. The graphs presented here underestimate the performance
improvement. 
\section{Concluding Remarks}
\label{sec-conclude}
Despite the various optimizations, 
with flooding-based routing, many 
routing messages are propagated unnecessarily.
%
We show that gossiping can reduce control traffic up to 35\% when compared
to flooding.
Since the routes found by gossiping may be up to 10-15\% longer than
those found by flooding (depending on the gossip probability),
how much gossiping can save in terms of overall traffic depends on
the gossip probability used, node
mobility, and the type of messages sent.
With high mobility, new routes will have to be found more frequently,
and the savings will be relatively greater. 
In addition, if messages are mainly network-wide broadcasts, rather than
point-to-point, gossiping may result in significant savings over
flooding.  (Note that with gossiping, in general, a small
fraction of the nodes will not get the broadcast.  However, in certain
application it may suffice that almost everyone gets the message, or the
contents of broadcast $k$ can be piggybacked with broadcast $k+1$, so
that the probability of missing a message altogether becomes very low.)

Our protocol is simple and easy to incorporate into
existing protocols. 
When we add gossiping to AODV,
simulations show significant performance
improvements 
in all the performance
metrics, even in networks 
as small as 150 nodes. 
As discussed in the
Section~\ref{sec-aodv}, we expect this performance improvement to become
even more significant in larger networks.

We have also experimented with adding gossiping to ZRP, by 
using gossiping to send the route request to 
some peripheral nodes 
rather than to all peripheral nodes.
Again, our results show significant improvement in all performance 
metrics.  
It seems likely
that gossiping can be usefully added to a number of other ad hoc
routing protocols as well.

Gossiping has a number of advantages over other approaches considered in
the literature.  For one thing,
unlike 
many heuristics considered in the literature, we believe that we
have a very good understanding of how gossiping will perform in large
networks.
This understanding is supported
both by analytical results and our experiments.
While there are fundamental limits to the amount of nonlocal traffic
that can be sent in large networks, due to problems of scaling
\cite{Gupta01,Jinyang01}, gossiping should still be useful in large
networks when nonlocal messages need to be sent.  
It is far less clear how well other optimizations
considered in the literature will perform in large networks.
Moreover, as our
simulations with AODV have shown, gossiping can provide significant
advantages even in small networks.
Experience in other contexts has shown
that gossiping is also quite robust and able to tolerate faults; we
expect that this will be the case in ad hoc routing as well.
All this suggests that gossiping can be a very useful adjunct to the
arsenal of techniques in mobile computing.  
Of course, work needs to be done in finding good techniques to learn
the appropriate gossip parameters. We have experimented with adjusting the
gossiping probability of each node according to the success/failure of
route requests; it is increased if the route request failure
probability is high, and decreased if the route request failure
probability is close to 0. To propagate the appropriate probability
throughout the network, it  can be put into the route request
packet. Each intermediate node receiving the packet will gossip with
the probability carried in the route request packet.
Our preliminary experiments have shown that this approach does
produce good results, 
 although we have not had enough experience 
 to determine the best way of making these adjustments to the gossip
 probability; we leave this for future work.

\subsection*{Acknowledgments} 
We would like to thank Jon Kleinberg for suggesting the relevance of
percolation theory, Harry Kesten for explaining the relevant results of
percolation theory, and Alan Demers for many useful comments.

\bibliographystyle{plain}
\bibliography{lilisbib}

\begin{thebibliography}{10}

\bibitem{DREAM98}
S.~Basagni, I.~Chlamtac, V.~R. Syrotiuk, and B.~A. Woodward.
\newblock A distance routing effect algorithm for mobility {(DREAM)}.
\newblock In {\em Proc.~Fourth Annual ACM/IEEE International Conference on
  Mobile Computing and Networking {(MobiCom)}}, pages 76--84, 1998.

\bibitem{Birman99}
K.~P. Birman, M.~Hayden, O.~Ozkasap, Z.~Xiao, M.~Budiu, and Y.~Minsky.
\newblock Bimodal multicast.
\newblock {\em ACM Transactions on Computer Systems}, 17(2):41--88, May 1999.

\bibitem{joch98}
J.~Broch, D.~A. Maltz, D.~B. Johnson, Y.~C. Hu, and J.~Jetcheva.
\newblock A performance comparison of multi-hop wireless ad hoc network routing
  protocols.
\newblock In {\em Proc.~Fourth Annual ACM/IEEE International Conference on
  Mobile Computing and Networking {(MobiCom)}}, pages 85--97, October 1998.

\bibitem{CRBirman01}
R.~Chandra, V.~Ramasubramanian, and K.~Birman.
\newblock Anonymous gossip: Improving multicast reliability in mobile ad-hoc
  networks.
\newblock In {\em Proc.~21st International Conference on Distributed Computing
  Systems {(ICDCS)}}, pages 275--283, 2001.

\bibitem{DasPerk00}
S.R. Das, C.~E. Perkins, and E.~M. Royer.
\newblock Performance comparison of two on-demand routing protocols for ad hoc
  networks.
\newblock In {\em Proc.~IEEE Conference on Computer Communications (INFOCOM)},
  pages 3--12, March 2000.

\bibitem{Demers87}
A.~Demers, D.~Greene, C.~Hauser, W.~Irish, J.~Larson, S.~Shenker, H.~Sturgis,
  D.~Swinehart, and D.~Terry.
\newblock Epidemic algorithms for replicated database maintenance.
\newblock In {\em Proc.~ACM Symposium on Principles of Distributed Computing},
  pages 1--12, 1987.

\bibitem{Grimmett89}
G.~Grimmett.
\newblock {\em Percolation}.
\newblock Springer-Verlag, New York,NY, 1989.

\bibitem{Gupta01}
P.~Gupta and P.~R. Kumar.
\newblock The capacity of wireless networks.
\newblock {\em IEEE Transactions on Information Theory}, IT-46(2):388--404,
  2000.

\bibitem{ZRP98}
Z.~Haas and M.~Pearlman.
\newblock The performance of query control schemes for the zone routing
  protocol.
\newblock In {\em Proc.~ACM SIGCOMM}, pages 167--177, August 1998.

\bibitem{Hari99}
W.~Heinzelman, J.~Kulik, and H.~Balakrishnan.
\newblock Adaptive protocols for information dissemination in wireless sensor
  networks.
\newblock In {\em Proc.~Fifth Annual ACM/IEEE International Conference on
  Mobile Computing and Networking {(MobiCom)}}, pages 174--185, 1999.

\bibitem{DSR96}
D.~B. Johnson and D.~A. Maltz.
\newblock {\em Dynamic Source Routing in Ad Hoc Wireless Networks}.
\newblock Kluwer Academic Publishers, 1996.

\bibitem{GPSR00}
B.~Karp and H.~T. Kung.
\newblock Greedy perimeter stateless routing {(GPSR)} for wireless networks.
\newblock In {\em Proc.~Sixth Annual ACM/IEEE International Conference on
  Mobile Computing and Networking {(MobiCom)}}, pages 243--254, 2000.

\bibitem{LAR98}
Y.~B. Ko and N.~H. Vaidya.
\newblock Location-aided routing {(LAR)} in mobile ad hoc networks.
\newblock In {\em Proc.~Fourth Annual ACM/IEEE International Conference on
  Mobile Computing and Networking {(MobiCom)}}, pages 66--75, 1998.

\bibitem{Jinyang01}
J.~Li, C.~Blake, D.~S.~J.~De Couto, H.~I. Lee, and R.~Morris.
\newblock Capacity of ad hoc wireless networks.
\newblock In {\em Proc.~seventh Annual ACM/IEEE International Conference on
  Mobile Computing and Networking {(MobiCom)}}, pages 61--69, 2001.

\bibitem{meester96}
R.~Meester and R.~Roy.
\newblock {\em Continuum percolation}.
\newblock Cambridge University, 1996.

\bibitem{NTCS99}
S.~Y. Ni, Y.~C. Tseng, Y.~S. Chen, and J.~P. Sheu.
\newblock The broadcast storm problem in a mobile ad hoc network.
\newblock In {\em Proc.~fifth Annual ACM/IEEE International Conference on
  Mobile Computing and Networking {(MobiCom)}}, pages 151--162, 1999.

\bibitem{TORA97}
V.~Park and M.S. Corson.
\newblock A highly adaptive distributed routing algorithm for mobile wireless
  networks.
\newblock In {\em Proc.~INFOCOM}, pages 1405--1413, April 1997.

\bibitem{AODV99}
C.~E. Perkins and E.~M. Royer.
\newblock Ad-hoc on-demand distance vector routing.
\newblock In {\em Proc.~2nd IEEE Workshop on Mobile Computing Systems and
  Applications}, pages 90--100, February 1999.

\bibitem{ns_2}
VINT Project.
\newblock The {UCB/LBNL/VINT} network simulator-ns ({V}ersion 2).
\newblock http://www.isi.edu/nsnam/ns.

\bibitem{rap96}
T.~S. Rappaport.
\newblock {\em Wireless Communications: Principles and Practice}.
\newblock Prentice Hall, 1996.

\bibitem{lucent-wavelan}
Bruce Tuch.
\newblock Development of {WaveLAN}, an {ISM} band wireless {LAN}.
\newblock {\em AT\&T Technical Journal}, 72(4):27--33, 1993.

\bibitem{Vahdat00}
A.~Vahdat and D.~Becker.
\newblock Epidemic routing for partially-connected ad hoc networks.
\newblock Duke Technical Report CS-2000-06, July 2000.

\end{thebibliography}

\end{document}